\documentclass{aa}  

\usepackage{graphicx}
\usepackage{txfonts}
\RequirePackage[normalem]{ulem} 
\RequirePackage{color}\definecolor{RED}{rgb}{1,0,0}\definecolor{BLUE}{rgb}{0,0,1} 
\providecommand{\DIFadd}[1]{{\protect\color{blue}\uwave{#1}}} 
\providecommand{\DIFdel}[1]{{\protect\color{red}\sout{#1}}}                      
\providecommand{\DIFaddbegin}{} 
\providecommand{\DIFaddend}{} 
\providecommand{\DIFdelbegin}{} 
\providecommand{\DIFdelend}{} 
\providecommand{\DIFaddbeginFL}{} 
\providecommand{\DIFaddendFL}{} 
\providecommand{\DIFdelbeginFL}{} 
\providecommand{\DIFdelendFL}{} 
\newcommand{\DIFscaledelfig}{0.5}
\RequirePackage{settobox} 
\RequirePackage{letltxmacro} 
\newsavebox{\DIFdelgraphicsbox} 
\newlength{\DIFdelgraphicswidth} 
\newlength{\DIFdelgraphicsheight} 
\LetLtxMacro{\DIFOincludegraphics}{\includegraphics} 
\newcommand{\DIFaddincludegraphics}[2][]{{\color{blue}\fbox{\DIFOincludegraphics[#1]{#2}}}} 
\newcommand{\DIFdelincludegraphics}[2][]{
        \sbox{\DIFdelgraphicsbox}{\DIFOincludegraphics[#1]{#2}}
        \settoboxwidth{\DIFdelgraphicswidth}{\DIFdelgraphicsbox} 
        \settoboxtotalheight{\DIFdelgraphicsheight}{\DIFdelgraphicsbox} 
        \scalebox{\DIFscaledelfig}{
                \parbox[b]{\DIFdelgraphicswidth}{\usebox{\DIFdelgraphicsbox}\\[-\baselineskip] \rule{\DIFdelgraphicswidth}{0em}}\llap{\resizebox{\DIFdelgraphicswidth}{\DIFdelgraphicsheight}{
                                \setlength{\unitlength}{\DIFdelgraphicswidth}
                                \begin{picture}(1,1)
                                \thicklines\linethickness{2pt} 
                                {\color[rgb]{1,0,0}\put(0,0){\framebox(1,1){}}}
                                {\color[rgb]{1,0,0}\put(0,0){\line( 1,1){1}}}
                                {\color[rgb]{1,0,0}\put(0,1){\line(1,-1){1}}}
                                \end{picture}
                        }\hspace*{3pt}}} 
} 
\LetLtxMacro{\DIFOaddbegin}{\DIFaddbegin} 
\LetLtxMacro{\DIFOaddend}{\DIFaddend} 
\LetLtxMacro{\DIFOdelbegin}{\DIFdelbegin} 
\LetLtxMacro{\DIFOdelend}{\DIFdelend} 
\DeclareRobustCommand{\DIFaddbegin}{\DIFOaddbegin \let\includegraphics\DIFaddincludegraphics} 
\DeclareRobustCommand{\DIFaddend}{\DIFOaddend \let\includegraphics\DIFOincludegraphics} 
\DeclareRobustCommand{\DIFdelbegin}{\DIFOdelbegin \let\includegraphics\DIFdelincludegraphics} 
\DeclareRobustCommand{\DIFdelend}{\DIFOaddend \let\includegraphics\DIFOincludegraphics} 
\LetLtxMacro{\DIFOaddbeginFL}{\DIFaddbeginFL} 
\LetLtxMacro{\DIFOaddendFL}{\DIFaddendFL} 
\LetLtxMacro{\DIFOdelbeginFL}{\DIFdelbeginFL} 
\LetLtxMacro{\DIFOdelendFL}{\DIFdelendFL} 
\DeclareRobustCommand{\DIFaddbeginFL}{\DIFOaddbeginFL \let\includegraphics\DIFaddincludegraphics} 
\DeclareRobustCommand{\DIFaddendFL}{\DIFOaddendFL \let\includegraphics\DIFOincludegraphics} 
\DeclareRobustCommand{\DIFdelbeginFL}{\DIFOdelbeginFL \let\includegraphics\DIFdelincludegraphics} 
\DeclareRobustCommand{\DIFdelendFL}{\DIFOaddendFL \let\includegraphics\DIFOincludegraphics} 

\begin{document}

   \title{STIX X-ray microflare observations during the Solar Orbiter commissioning phase}

  \author{Andrea Francesco Battaglia\inst{1,10} \and 
    Jonas Saqri\inst{2} \and
    Paolo Massa\inst{3} \and 
    Emma Perracchione\inst{3} \and  
    Ewan C. M. Dickson\inst{1,2} \and   
    Hualin Xiao\inst{1} \and
    Astrid M. Veronig\inst{2} \and     
        Alexander Warmuth\inst{4} \and        
    Marina Battaglia\inst{1} \and 
    Gordon J. Hurford\inst{1} \and    
    Aline Meuris\inst{5} \and     
        Olivier Limousin\inst{5} \and  
        László Etesi\inst{1} \and
    Shane A. Maloney\inst{6,7} \and     
    Richard A. Schwartz\inst{2,6,8,9,}\thanks{Deceased.} \and  
    Matej Kuhar\inst{1} \and   
        Frederic Schuller\inst{4} \and  
    Valliappan Senthamizh Pavai\inst{4} \and
    Sophie Musset\inst{11} \and
    Daniel F. Ryan\inst{1} \and
        Lucia Kleint\inst{12} \and       
    Michele Piana\inst{3} \and     
        Anna Maria Massone\inst{3} \and    
        Federico Benvenuto\inst{3} \and
        Janusz Sylwester\inst{13} \and     
        Michalina Litwicka\inst{13} \and
    Marek Stęślicki\inst{13} \and     
        Tomasz Mrozek\inst{13,14} \and 
        Nicole Vilmer\inst{15} \and     
        Franti\v{s}ek F\'{a}rn\'{\i}k\inst{16} \and     
        Jana Ka\v{s}parov\'{a}\inst{16} \and    
        Gottfried Mann\inst{4} \and     
        Peter T. Gallagher\inst{6,7} \and  
        Brian R. Dennis\inst{9} \and
    André Csillaghy\inst{1} \and
    Arnold O. Benz\inst{1} \and
    S\"am Krucker\inst{1,17}
     }

   \institute{
             University of Applied Sciences and Arts Northwestern Switzerland, Bahnhofstrasse 6, 5210 Windisch, Switzerland 
             \and
                 Institute of Physics, University of Graz, A-8010 Graz, Austria
                     \and 
             Dipartimento di Matematica, Università di Genova, via Dodecaneso 35, I-16146 Genova, Italy
             \and      
             Leibniz-Institut f\"ur Astrophysik Potsdam (AIP), An der Sternwarte 16, D-14482 Potsdam, Germany
             \and
                 AIM, CEA, CNRS, Université Paris-Saclay, Université Paris Diderot, Sorbonne Paris Cité, F-91191 Gif-sur-Yvette, France 
                 \and
                 Astrophysics Research Group, School of Physics, Trinity College Dublin, Dublin 2, Ireland
                 \and 
             School of Cosmic Physics, Dublin Institute for Advanced Studies, 31 Fitzwilliam Place, Dublin, D02 XF86, Ireland
             \and
             American University, 4400 Massachusetts Ave, NW, Washington DC 20016, USA
                 \and
                 Solar Physics Laboratory, Code 671, NASA Goddard Space Flight Center, Greenbelt, MD, USA
                 \and
             ETH Z\"urich, R\"amistrasse 101, 8092 Z\"urich, Switzerland 
             \and
                 SUPA, School of Physics and Astronomy, University of Glasgow, Glasgow G12 8QQ, UK
                 \and
             University of Geneva, CUI, 1227 Carouge, Switzerland
             \and    
             Space Research Centre, Polish Academy of Sciences, Bartycka 18A, 00-716 Warszawa, Poland
                 \and
             Astronomical Institute, University of Wroclaw, Wroclaw, Poland
                 \and
             LESIA, Observatoire de Paris, Université PSL, CNRS, Sorbonne Université, Université de Paris, 5 place Jules Janssen, 92195 Meudon, France
                 \and
                 Astronomical Institute of the Czech Academy of Sciences, Fri\v{c}ova 298, Ond\v{r}ejov, Czech Republic
                 \and
             Space Sciences Laboratory, University of California, 7 Gauss Way, 94720 Berkeley, USA
}
\authorrunning{A. F. Battaglia et~al.}

   \date{Received 10 February 2021 / Accepted 31 May 2021}

 
  \abstract
   {
   The Spectrometer/Telescope for Imaging X-rays (STIX) is the hard X-ray instrument onboard Solar Orbiter designed to observe solar flares over a broad range of flare sizes.
   }
   {
   We report the first STIX observations of solar microflares recorded during the instrument commissioning phase in order to investigate the STIX performance at its detection limit.
   }
   {
   STIX uses hard X-ray imaging spectroscopy in the range between 4-150 keV to diagnose the hottest flare plasma and related nonthermal electrons. This first result paper focuses on the temporal and spectral evolution of STIX microflares occuring in the Active Region (AR) AR12765 in June 2020, and compares the STIX measurements with Earth-orbiting observatories such as the X-ray Sensor of the  Geostationary Operational Environmental Satellite (GOES/XRS), the Atmospheric Imaging Assembly of the Solar Dynamics Observatory (SDO/AIA), and the X-ray Telescope of the Hinode mission (Hinode/XRT).
   }
   {
   For the observed microflares of the GOES A and B class, the STIX peak time at lowest energies is located in the impulsive phase of the flares, well before the GOES peak time. Such a behavior can either be explained by the higher sensitivity of STIX to higher temperatures compared to GOES, or due to the existence of a nonthermal component reaching down to low energies. The interpretation is inconclusive due to limited counting statistics for all but the largest flare in our sample. For this largest flare, the low-energy peak time is clearly due to thermal emission, and the nonthermal component seen at higher energies occurs even earlier. This suggests that the classic thermal explanation might also be favored for the majority of the smaller flares.  In combination with EUV and soft X-ray observations, STIX corroborates earlier findings that an isothermal assumption is of limited validity. Future diagnostic efforts should focus on multi-wavelength studies to derive differential emission measure distributions over a wide range of temperatures to accurately describe the energetics of solar flares.
   }
   {
   Commissioning observations confirm that STIX is working as designed. As a rule of thumb, STIX detects flares as small as the GOES A class. For flares above the GOES B class, detailed spectral and imaging analyses can be performed.
   }
   \maketitle
   
\section{Introduction}

Microflares are dynamic, small-scale energy release events in the solar atmosphere. They are presumably driven by the same physical processes as larger flares, but with the energy release by magnetic reconnection being many orders of magnitude smaller. However, their occurrence frequency is much higher and thus they may play an important role in heating of the solar corona and supplying its mass \citep[e.g.,][]{1991SoPh..133..357H}. It has been shown that the occurrence frequencies and energy distributions from the smallest nano- (quiet Sun) and microflares to the largest X-class flares follow a power-law, with the power-law index $\alpha$ obtained from different studies, instruments, and wavelengths typically in the range of $1.5 \lesssim \alpha \lesssim 2.5$
\citep{1993SoPh..143..275C,1995PASJ...47..251S,1998ApJ...501L.213K,2000ApJ...529..554P,2002A&A...382.1070V,2008ApJ...677.1385C,2008ApJ...677..704H,2011SSRv..159..263H}. Observations at X-ray wavelengths provide us with the most direct insight into the energy release and energy conversion in solar (micro-)flares, as they allow us to diagnose the properties of the accelerated electrons, the heating of the flaring plasma, and part of the atmospheric response to the energy input by the electron beams. Since the latter can carry a substantial amount of the total flare energy \citep[e.g.][]{2003AdSpR..32.2459D,2012ApJ...759...71E,2016ApJ...832...27A,2020A&A...644A.172W}, it is essential to accurately determine their total energy. 

The first hard X-ray (HXR) microflare observations by a balloon-borne experiment \citep{1983BAAS...15..712S,1984ApJ...283..421L} have already shown that in microflares 
(as in larger flares), the HXR spectra may reveal a power-law component indicative of nonthermal bremsstrahlung from electron beams
\citep{1971SoPh...18..489B,1971SoPh...17..412L}. The Reuven Ramaty High Energy Solar Spectroscopic Imager \citep[RHESSI, in operation 2002--2018;][]{2002SoPh..210....3L} has so far provided us with the largest observational data base of microflares at HXR wavelengths. The nonthermal photon power-law indices $\gamma$ in microflares are typically larger than in larger flares, with  $4 \lesssim \gamma \lesssim 10$ above energies of $\approx$10 keV, indicative of a softer spectrum of the accelerated electrons \citep{2002SoPh..210..445K,2005A&A...439..737B,2007SoPh..246..339S,2008ApJ...677.1385C,2008ApJ...677..704H,2016A&A...588A.115W}. However, some microflares revealed HXR photon spectra as hard as $\gamma \approx 2.5$ down to energies as low as 4 keV \citep{2008A&A...481L..45H}, and occasionally extend above 100 keV \citep{2013ApJ...765..143I}. How low in energy the nonthermal component extends is an essential diagnostic, as most of the energy in nonthermal electrons for these steep spectra resides in the low energy end. However, this low energy cutoff is notoriously difficult to infer observationally \citep[e.g.][]{2019ApJ...881....1A,2019ApJ...871..225K}. Recent high-sensitivity observations with the Nuclear Spectroscopic Telescope Array \citep[NuSTAR,][]{2013ApJ...770..103H} showed that in at least in 1 out of 11 microflares, there is a clear indication that the nonthermal emission dominates the count rates down to $<$5 keV \citep{2020ApJ...891L..34G,2020arXiv201106651D}. 

Multi-wavelength observations of RHESSI microflares have revealed common features as well as a significant variety and complexity of their appearances. As in the case of larger flares, HXR microflares occur in Active Regions \citep[AR;][]{2007SoPh..246..339S,2008ApJ...677.1385C}. Contrary to what their name may suggest, they are not necessarily spatially small \citep{2008ApJ...677..704H}. At lower HXR energies, RHESSI microflares typically show an elongated structure, indicative of a flare loop. The higher-energy HXR emission tends to be concentrated at footpoints of the loops, which are rooted in opposite magnetic polarity regions \citep{2002SoPh..210..445K,2004ApJ...604..442L,2007SoPh..246..339S,2009A&A...505..811B,2011SSRv..159..263H}. Despite the ``single loop'' structure suggested by the HXR imaging, the coordinated high-resolution imagery and spectroscopy at EUV and optical wavelengths revealed more complex fine structure and very dynamic responses of the transition region and chromosphere to the microflare energy input with simultaneous upward and downward directed plasma flows due to chromospheric evaporation \citep{2008SoPh..250..315S,2009ApJ...692..492B,2009A&A...505..811B,2017ApJ...845..122G, 2019ApJ...881..109H,2020ApJ...891...78A,2020arXiv201104753V}. A number of RHESSI microflares have been reported to be associated with EUV jets and radio emission, such as type III radio bursts \citep{2004ApJ...604..442L,2007SoPh..246..339S,2009A&A...505..811B,2018ApJ...866...62C,2012ApJ...754....9G,2020ApJ...889..183M,2020ApJ...904...94S}, indicative of a multipolar magnetic field topology involved in the energy release process, and electron beams being accelerated along closed loops as well as along field lines that are ``open'' to interplanetary space.

With the launch and commissioning of the Solar Orbiter mission, the Spectrometer/Telescope for Imaging X-rays \citep[STIX;][]{2020A&A...642A..15K} is the latest HXR telescope to study solar flares. While the HXR diagnostics capabilities of STIX resemble its predecessor, RHESSI, its unique orbit away from the Earth-Sun line in combination with the opportunity of joint observations with other Solar Orbiter instruments will provide essential new inputs into understanding the magnetic energy release and particle acceleration in solar flares. This paper presents observations of microflares taken during the commissioning phase in June 2020 to demonstrate STIX diagnostics capabilities at its sensitivity limit and to describe the STIX data products beyond what is mentioned in the STIX instrument paper  \citep{2020A&A...642A..15K}. Additionally, we shed new light on the discussion of the low energy extension of nonthermal emission to better constrain the energy content in accelerated electrons in microflares. 

The paper is structured as follows. In Sect.\,2, we present a simulation of the STIX instrument response to the X-ray flare spectra in the context of the diagnostic capabilities of STIX with a focus on microflares. In Sect.\,3, the newly available observations are first discussed on a statistical basis comparing STIX and GOES microflares, followed by a more detailed discussion of three individual microflares, where we consider additional observations by SDO/AIA and HINODE/XRT of the microflare's spatial and temporal evolution. Finally, we give a summary of our findings in Sect.\,~\ref{sec:summary}.

\section{STIX microflare diagnostics}

\begin{figure}[tbp]
\centering 
\includegraphics[height=0.8\textwidth,keepaspectratio,trim=260 70 260 70, clip]{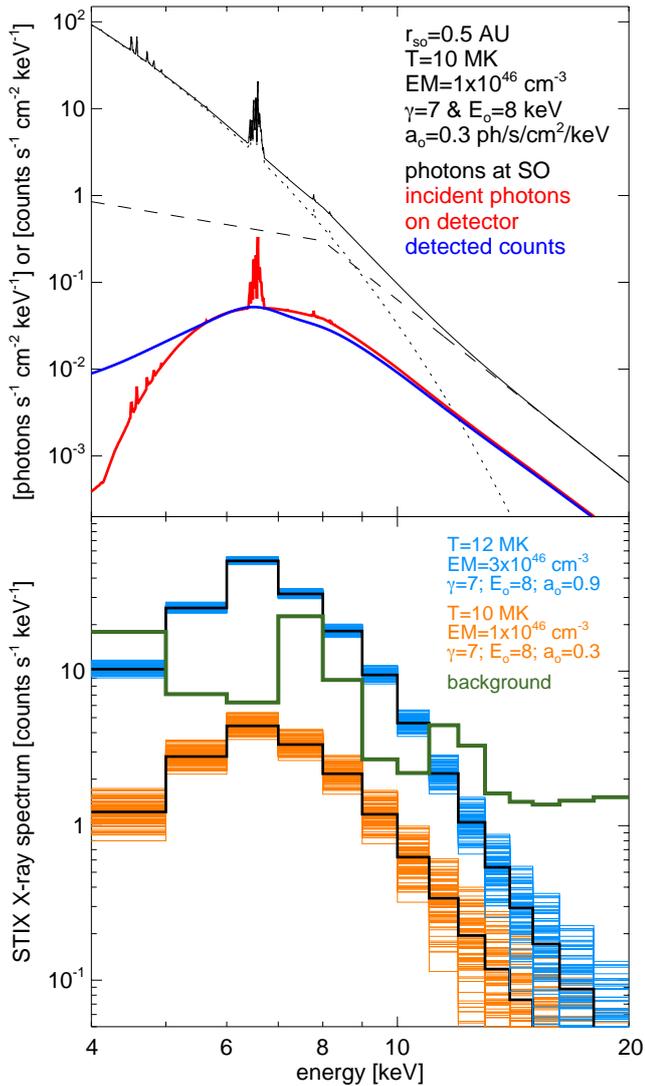}
\caption{Simulated STIX instrument response to microflares by assuming an isothermal optically thin source using coronal abundances plus an additional nonthermal tail with a power-law shape and a break energy of the photon spectrum. The simulations were performed for a radial distance from the Sun of  0.5 AU and a flare location at $\mathrm{S}800^{\prime\prime}~\mathrm{E}1600^{\prime\prime}$
(these parameters roughly correspond to the observations shown in Sect.\,3). 
The top panel shows the arriving photon spectrum at
Solar Orbiter (black, with the thermal and nonthermal components shown in dotted and dashed, respectively), the attenuated photon spectrum incident on the STIX detectors (red), and the resulting count spectrum (blue). The bottom panel shows the count spectrum after onboard binning into the STIX science energy channels. The light blue and the orange curves are simulations for two different microflares as given in the annotations. The different histograms in the same color represent different realizations of the counting statistics assuming an integration time of 30 s. The black histograms are the simulations without the noise from counting statistics. The green histogram is the instrumental background that is stable in time and can therefore be measured much more accurately than the microflare count spectra.}
\label{Fig:sim}
\end{figure}

STIX is designed to observe a wide range of solar flares. To do so, high count statistics for the largest solar flares have to be attenuated in order to avoid detector saturation. A $0.6$ mm thick aluminum attenuator, placed into the detector field of view, is inserted and removed autonomously on the decision of an onboard Rate Control Regime algorithm \citep[RCR;][]{2020A&A...642A..15K}. On the other hand, while the RCR handles the count rates of the largest solar flares, the limits for the detection of the smallest flares come from the instrumental background produced by the onboard radioactive calibration source (Ba-133). The main emission lines used to calibrate STIX are at 31 keV and 80 keV. However, Ba-133 also produces CdTe escape lines at the lowest energies of the STIX range (see Fig.\,\ref{Fig:sim}, bottom panel, green curve). These lines affect all STIX science low-energy bins with the ones from 5 to 7 keV and above 9 keV being least affected. Below $\sim$18 keV, the instrumental background is much stronger than the cosmic X-ray emission that passes through the STIX imaging grid apertures. As the half life time of Ba-133 is 10.7 years, the background is very stable in time making background subtraction feasible even for microflares with count rates well below the instrumental background rate.

To demonstrate the STIX sensitivity to microflares, we modelled the STIX response to an isothermal flare with a nonthermal component using the current (October 2020) best STIX response model. The top panel in Fig.\,\ref{Fig:sim} shows the arriving X-ray photon spectrum for a flare with a temperature of 10 MK, an emission measure (EM) of $1\times10^{46}$ cm$^{-3}$, assuming coronal abundances, and a nonthermal power-law spectrum with a slope of $\gamma = 7$, a break energy of the photon spectrum at $E_0 = 8\,\mathrm{keV}$ and a normalization factor at the break energy  $a_0=0.3\,\mathrm{ph}\,\mathrm{s}^{-1}\,\mathrm{cm}^{-2}\,\mathrm{keV}^{-1}$. Below the break, the photon spectrum is assumed to be flat with a slope of 1.5 to simulate the effect of a cutoff energy in the electron spectrum. The incoming photon spectrum is then attenuated by various entrance windows within the instrument before photons arrive at the detectors (i.e., the STIX entrance windows, the STIX imaging grids, the various multi-layer insulations, and the detector dead layer\footnote{The transmission is still under evaluation using the STIX crab nebula observations. First results show that the current calibration is better than 20\% accuracy, near the typical value of the absolute calibration uncertainty between different X-ray observatories \citep{2005SPIE.5898...22K,2015ApJS..220....8M}}) resulting in an incident photon spectrum that is heavily attenuated at the lowest energies (see red curve in the top panel of Fig.\,1). At higher energies, the attenuation is mainly due to the STIX imaging system (tungsten grids) that attenuates the signal by roughly a factor of 4 (i.e., half the photons go through each grid). The detector response matrix is then used to calculate the count spectrum that results from the incoming attenuated photon spectrum. With the 1 keV resolution of the STIX detectors, the high-energy part of the count spectrum has a similar shape to the photon spectrum. At low energies, however, the spectrum is smeared out by the non-diagonal elements of the response matrix. The individual lines in the Fe complex are therefore no longer resolvable. We also note that the lowest STIX science energy channel from 4 to 5 keV is predominantly produced by photons at higher energies. This is due to the very strongly increasing attenuation of the incoming photons for decreasing photon energies \citep[see Fig. 3 of][]{2020A&A...642A..15K}. For the lowest STIX energy channel (4-5 keV), the attenuation falls off steeper compared to the smearing out of 
Hence, the recorded 4-5 keV counts are mainly produced by photons at higher energies for which the detector collects only a fraction of photon's energy. As same photons are registered at a too low energies, the count spectrum (blue) is consequently slightly below the incoming photon spectrum (red) for energies around the peak in the count spectrum. While the 4-5 keV counts do not reflect the actual photons in that energy, they still carry spectral information and they are therefore nevertheless valuable diagnostics in spectral fitting.
The 4-5 keV channels is also important for the detection of cooler X-rays sources, such as quiescent active regions for which the photon spectrum is steeper and in relative terms there are more 4 to 5 keV photons arriving at the detectors compared the photons with energies around the peak count spectrum (i.e., 6-7 keV here).

To cope with the limited telemetry, the STIX data are binned onboard in time and energy \citep[see][]{2020A&A...642A..15K}. This is in contrast to previous HXR missions such as RHESSI \citep{2002SoPh..210....3L}, where each individual photon was time and energy tagged. However, this allows us to efficiently use the telemetry allocated to STIX. There are 32 STIX science energy bins which are currently selected to have a width of 1 keV at the lowest energies. The bottom panel in Fig.\,1 gives the STIX count spectrum of the simulated microflare for 100 different representations of the counting statistics assuming a 30 s integration time (orange curves). The peak in the count spectrum is in the 6-7 keV bin and the count spectrum of the microflare alone lies below the instrumental background (green curve) for all energies. As the background is stable in time and can be measured during non-flaring times, even small events such as modelled here can clearly be measured by STIX. 
The light blue curves show the result of a slightly larger and hotter microflare ($3\times10^{46}$ cm$^{-3}$ and 12 MK). For this slightly stronger microflare, the expected count rates are already above the instrumental background for energies between 5 and 11 keV. For such a microflare, a nonthermal component can indeed be measured. 

While it is, in principle, straightforward to calculate the GOES class for the simulated flares, which correspond to A2 and A7, respectively, this should not be interpreted as determining the likely GOES class associated with these two STIX simulations. As GOES observations reach down to lower X-ray energies than STIX, GOES fluxes can additionally contain emissions from cooler flare plasma than STIX. As flares  are generally multi-thermal in nature, the associated GOES class will therefore be higher than the estimate obtained by using just the emission that STIX detects \citep[e.g.,][]{2009ApJ...697...94M, 2014SoPh..289.2547R, 2016A&A...588A.115W}. As the difference depends on the shape of the differential emission measure distribution, an associated GOES class cannot be readily estimated. As typical values, we propose that the potential GOES class associated with these STIX simulations are possibly a factor of 2 to 4 larger than estimated above (e.g., A4 to A8 (orange) and B1 to B3 (blue), respectively).

\section{Observations}

\begin{figure*}[tbp]
\centering 
\includegraphics[height=0.44\textwidth,keepaspectratio,trim=20 245 38 0, clip]{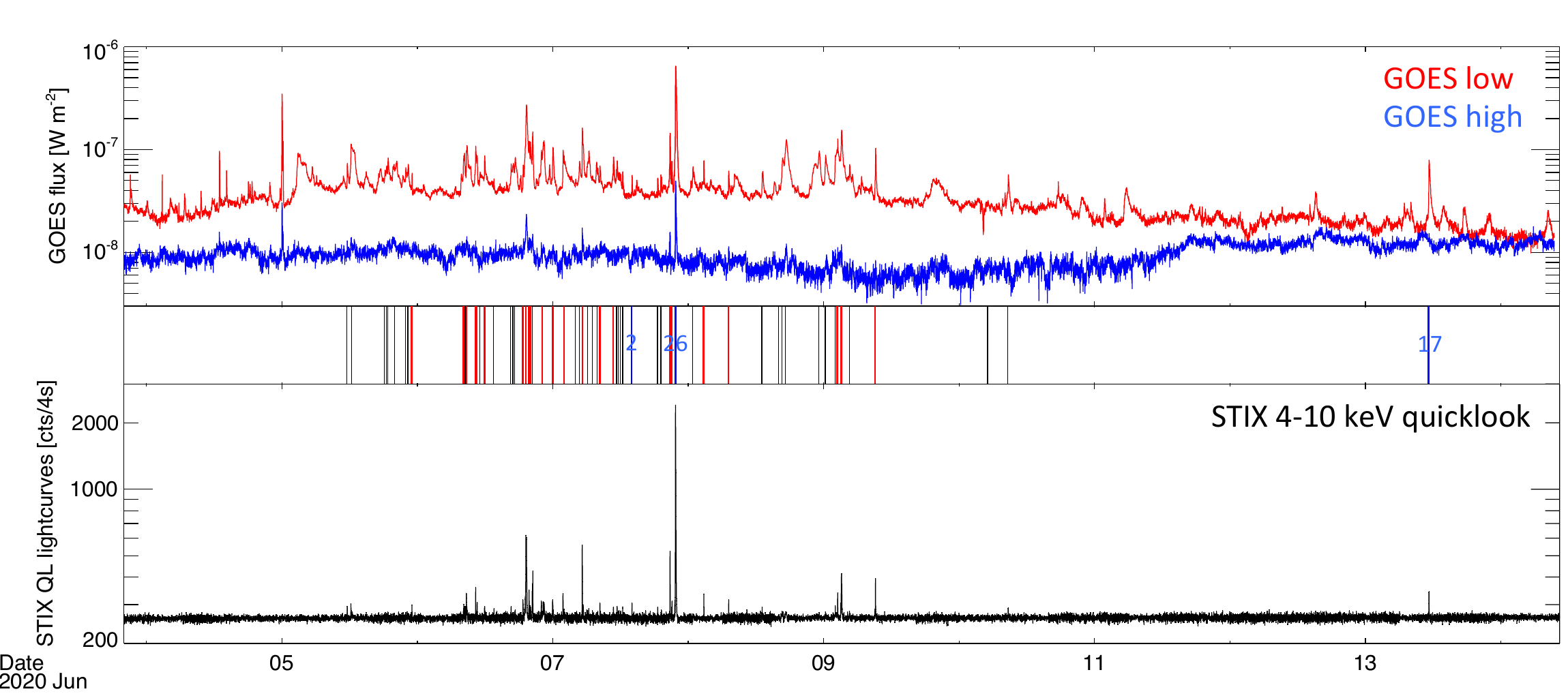}
\caption{GOES Soft X-ray profiles (\emph{top}) and STIX 4-10 keV quicklook light curve (\emph{bottom}) during the time period considered in the paper. The STIX quicklook light curve was used to select 69 flares, which are marked by vertical bars in the central panel. Most of these events were only seen in a single STIX science energy channel (\emph{middle, black}), and they are therefore difficult to see in the displayed 4-10 keV light curve. For the 26 events which are at multiple STIX science energy bins, flare temperature could be derived (\emph{middle, red}). The three flares discussed in detail are marked and numbered in blue. We note that all flares are from AR12765, which only became visible from Solar Orbiter after June 5. Hence, flares from earlier times were occulted by the solar limb and not visible by STIX. }
\label{Fig:qllc}
\end{figure*}

Despite the fact that the commissioning phase was intended for instrument checkout and calibration, the acquired data is generally taken at full instrument performance and can therefore be used for scientific studies. The best observing conditions for microflares during the Solar Orbiter commissioning phase were between June 5 and June 21, 2020, after active region AR12765 appeared on disk from the Solar Orbiter vantage point. During this period, Solar Orbiter was roughly at 0.52 AU from the Sun and between 41 and 74 degrees west of the Earth-Sun line. As these observations were taken during a commissioning phase outside the nominal science phase of Solar Orbiter, no coverage from the other Solar Orbiter instruments is available during this time period. However, AR12765 was also on-disk as seen from Earth and all microflares presented in this paper were therefore visible to several Earth-orbiting observatories.

To best allocate its limited telemetry, STIX first sends to the ground low resolution QuickLook (QL) data \citep[for details see][]{2020A&A...642A..15K}. Figure\,\ref{Fig:qllc} shows the STIX QL light curve at the lowest energy channel (4-10 keV) during the time period from June 4 through June 14. From the QL data, the STIX team then selects the most scientifically promising flares, and so-called data requests are sent to STIX to download the selected flares at the chosen temporal and spectral resolution. From the QL data during commissioning, a total of 69 microflares were selected with the majority of events (53 of 69) having a background subtracted count spectrum below the instrumental background rate. Many microflares were only seen in a single STIX science energy channel with rather low counting statistics. To take advantage of STIX temperature diagnostic capabilities, we restricted the statistical analysis to the 26 microflares, which are clearly seen in at least two energy channels (see Sect.\,\ref{sec:statistical_analysis}), while for three events, we present a more in-depth analysis at different wavelengths in Sect.\,\ref{sec:multi-wavelength_analysis}. Prior to the discussion of the scientific findings, the data analysis approach is outlined in the following two sections for GOES and STIX individually. 

\subsection{GOES \label{sec:GOESdata}}

The GOES analysis was done using GOES-16 observations and the latest version of the GOES Workbench within SSWIDL (status October 2020). This software package applies a correction factor to GOES to match the previously used GOES flare classification. For all flares, an individual GOES background has been subtracted depending on the time variations observed before and after the selected events. For many cases, a constant pre-flare background subtraction gives good results. For flares occurring during the decay of a previous event, a time dependent background was selected. As all values of the GOES flare class given in this paper are from background subtracted light curves, they might be somewhat smaller than the values given in the official flare catalogues, which do not have backgrounds subtracted. To reduce the noise in the time profiles, the GOES curves shown in this paper have been integrated from the nominal 1 s resolution to 15 s time bins. This reduction in time resolution is appropriate considering typical rise and decay times of microflares of a minute or longer. The time derivative of the GOES 1--8 {\AA} low energy channel is used as a proxy to roughly outline the duration of the impulsive phase of the flare. As the derivative can be very noisy, a box car smoothing with a window of 30 or 45 s has been applied for smaller events before calculating the derivative. GOES temperature estimates were done using the same standard software, but using integration times of one minute in order to further mitigate noise issues. To double check, the temperature estimates were additionally derived from GOES-17 data.

\subsection{STIX \label{sec:STIXdata}}
 
To maximize the counting statistics in our sample of microflares, our analysis mainly relies on the averaged count rates over the 30 imaging detectors. There are two STIX data products that can be used to get detector-averaged counts. The first product is the L1 'pixel' data, which contains information on individual detectors and pixels, with the cost of using up telemetry faster. With this type of data product, the detector-averaging is done on the ground, during data analysis. On the other hand, in order to save telemetry, the detector-averaging can be done onboard by the flight-software: this is the case for the L4 'spectrogram' data, the second STIX data product to get detector-averaged counts \citep{2020A&A...642A..15K}. The analysis presented in this paper used L4 spectrogram data for all events except otherwise noted, and the total requested data for all 69 flares is only 0.3 MiB. In the following, we briefly summarize the different aspects of data analysis that have been considered:

\begin{itemize}
    \item \textbf{Attenuator:} The STIX attenuator was out providing maximal sensitivity at the lowest energies. The STIX attenuator is designed to only be activated for $\sim$M class flares. As a side note, we mention here that with the attenuator inserted, only the largest event in our sample would have been detected.
    \item \textbf{Background:} The STIX background below $\sim$18 keV is dominated by counts related to the internal calibration source and therefore the background is stable in time. As the detector calibration depends very slightly on detector temperature, the recorded background also varies slightly with detector temperature. However, temperature effects are greatly suppressed by the STIX detector system by using a baseline holder feature \citep[BSH;][]{2021JAI....1050009G}. Temperature variations are mainly driven by the cyclic passive cooling provided by the spacecraft radiatiors to STIX. For observations at 0.52~AU, the radiator temperature still hits the low temperature limit of $-40^{\circ}$C and heaters are therefore periodically activated. This results in temperature variations from $-40^{\circ}$C to $-38^{\circ}$C and back to $-40^{\circ}$C every $\sim$40 minutes, causing a slight shift in energy in the detector calibration. Consequently the background spectrum is minimally shifted relative to the fixed energy bins. The largest effects of the temperature dependence is seen in energy bins that are dominated by the wings of an emission line of the background spectrum. 
    The maximal effect is seen in the 8-9 keV science bin at 3\% increase per $^{\circ}$C, and correspondingly at 6-7 keV with a 2\% decrease per $^{\circ}$C. Hence, the effect is rather small, and therefore only visible for the smaller flares in our sample. As the detector temperature is measured at one minute cadence, this effect can be corrected. We used observations during non-flaring times to measure the temperature dependence of the background and correct for it when we subtract the background. However, the results presented in this paper do not change even if this correction is not applied.   
    \item \textbf{Light travel time:} For all plots shown in this paper, the STIX times have been corrected for the different light travel time of Solar Orbiter, namely, the photon arrival time, relative to Earth. The correction ranges from 232.6 s on June 5 to 241.4 s on June 14. 
    \item \textbf{Time resolution:} The time resolution during commissioning was not yet optimized to save onboard memory, and it was set to a rather high cadence, resulting in time bins between 1 and 2 s, depending on the actual observed count rates. For the analysis presented here, the data has been integrated on the ground to 30 s or 90 s time bins, depending on the counting statistics. Meanwhile, the STIX cadence during nonflaring times has been set to ~20 s to optimize onboard memory usage. This setting will fill the onboard memory significantly more slowly and still allows us to get the same microflare diagnostics as presented in this work.
    \item \textbf{Errors in light curves:} The main contribution for the error bars shown in this paper is counting statistics. Errors due to the data compression are added, but are much smaller for these low-counting statistics observations (the correction for compression is only 0.25\% relative error even for the largest event in this sample). 
   \item \textbf{Livetime:} The detector livetime for these very small events is very close to 1. During non-flaring time, the livetime is 99.87\%, and decreases to 99.75\% for the largest event in our sample. The shown time profiles have been corrected for livetime, but the effect is well below the statistical errors and could in principle be neglected for the small events discussed here.
     \item \textbf{Energy resolution:} STIX bins counts into predefined science energy bins. For this analysis we used the nominal bin size, which results in $\sim$1 keV bin width for the energy ranges used here. The nominal 1 keV resolution is achieved by using passively cooled Cadmium Telluride (CdTe) X-ray detectors \citep{2020A&A...642A..15K}
    \item \textbf{Energy Look Up Table (ELUT):} As the energy calibration that is used onboard to bin the counts into the STIX science energy bins is different for each pixel, the actual bin width is slightly different for each pixel. The different bin width can be corrected during data analysis for L1 'pixel' data. For the L4 'spectrogram' data, the effect is smeared out. As a correction is well below the statistical errors for our set of flares, in this work, for simplicity, it is only applied during spectral fitting.
    \item \textbf{Transmission correction:} The light curves shown in this paper are not corrected for the transmission and therefore reflect count rates. The transmission correction is only applied to perform spectral fitting (see Sect.\,\ref{sec:spectroscopy}). As mentioned in Sect.\,2, the applied transmission corrections use current knowledge as of October 2020.
    \item \textbf{Temperature diagnostics:} For the 26 flares that were detected in at least two STIX science energy bands, temperature and emission measure estimates were performed, assuming an isothermal plasma. To have the ability to use the same method for all events, the ratio of the 6-7 keV to the 5-6 keV channel has been used as a temperature diagnostic. The simulations shown in Fig.\,\ref{Fig:sim} were repeated for an isothermal plasma (without adding a nonthermal component) for an array of temperatures taking the transmission function into account. These simulations give the expected STIX count rates as a function of temperature for each energy bin, and hence, the temperature dependence of the ratio of different energy bins. From the observed ratio and its uncertainty, a unique temperature and emission can be estimated with error bars. We note here that we will have software available in the near future within the SSWIDL OSPEX package that will properly fit the entire count spectrum making the currently used ratio method obsolete. For the GOES B6 flare occurred on June 7 (see Sect.\,\ref{sec:statistical_analysis}), a preliminary spectral fitting is performed by using for the first time the OSPEX SSWIDL package with STIX data: This is shown in the dedicated Sect.\,\ref{sec:spectroscopy}.
    \item \textbf{Imaging:} This first results STIX paper focuses on the time evolution and simple spectroscopy of microflares observed during commissioning. The current status of the STIX imaging calibration does not allow us yet to make images routinely. In any case, imaging information for most microflares presented in this paper will be limited, and it is only for the few largest flares in our sample with high enough statistics in individual detector pixels that imaging can provide scientifically meaningful information. However, one preliminary step toward STIX imaging is reported in Sect.\,3.4.4.
    \end{itemize}
In the following subsections, we first present the statistical results of the selected 26 microflares using STIX and GOES data only, followed by a more in-depth discussion of three microflares, where we also include EUV observations from AIA and soft X-ray observations by XRT. 
    
\subsection{Statistical analysis of the temporal evolution \label{sec:statistical_analysis}}

\begin{figure*}[tbp]
\centering 
\includegraphics[height=1.25\textwidth,keepaspectratio,trim=15 20 290 00, clip]{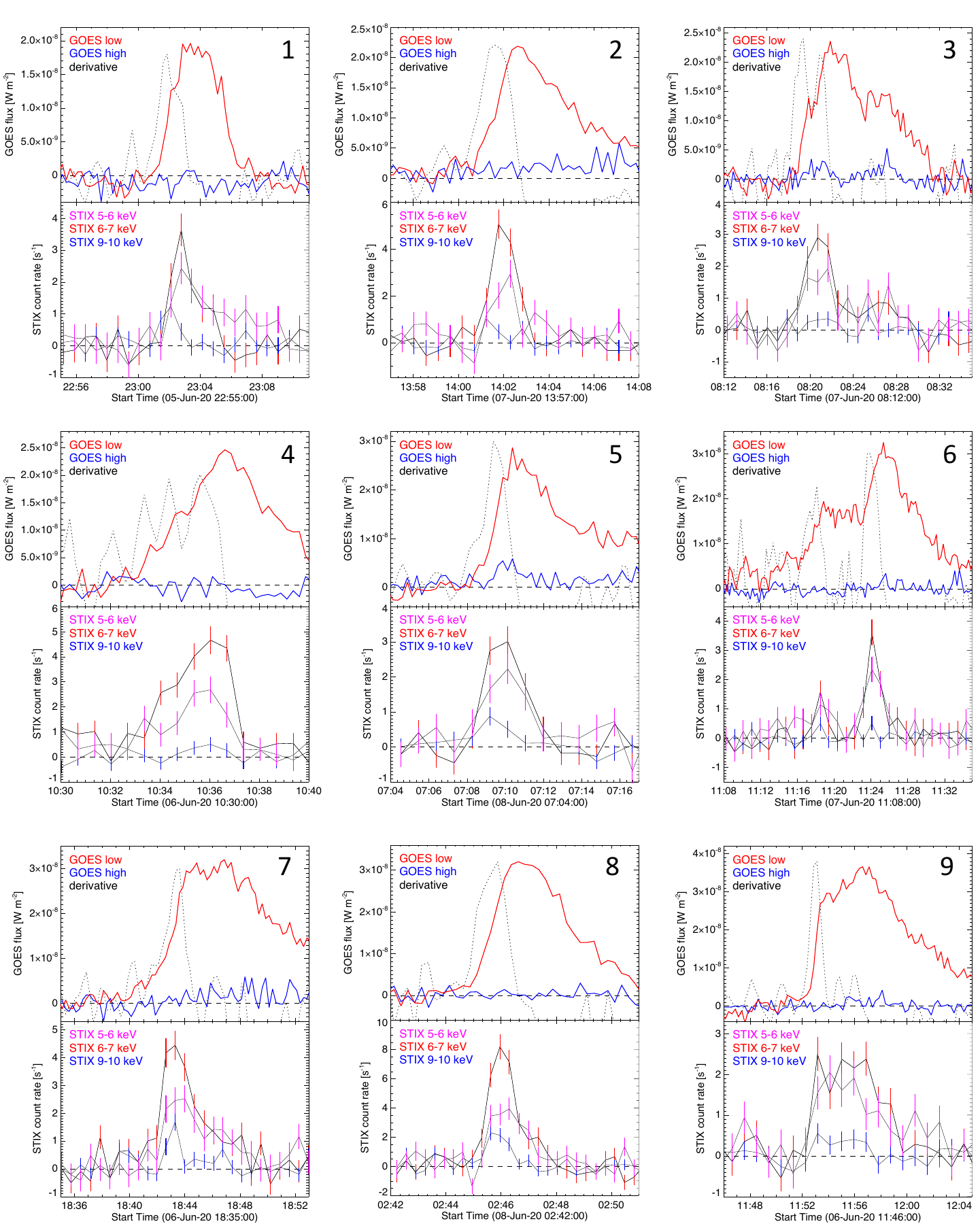}
\caption{GOES and STIX light curves for the smallest of the selected microflares of GOES classes below A4. For each flare, in the top panels the GOES light curves are shown after background subtraction, while below the background subtracted and livetime-corrected STIX count rates for three science energy bins are displayed with error bars. The STIX time profiles have been adjusted for the light travel time to Earth. For reference, the GOES low energy channel derivative is shown in dotted black.  }
\label{Fig:lc_small}
\end{figure*}

\begin{figure*}[tbp]
\centering 
\includegraphics[height=1.25\textwidth,keepaspectratio,trim=15 20 290 00, clip]{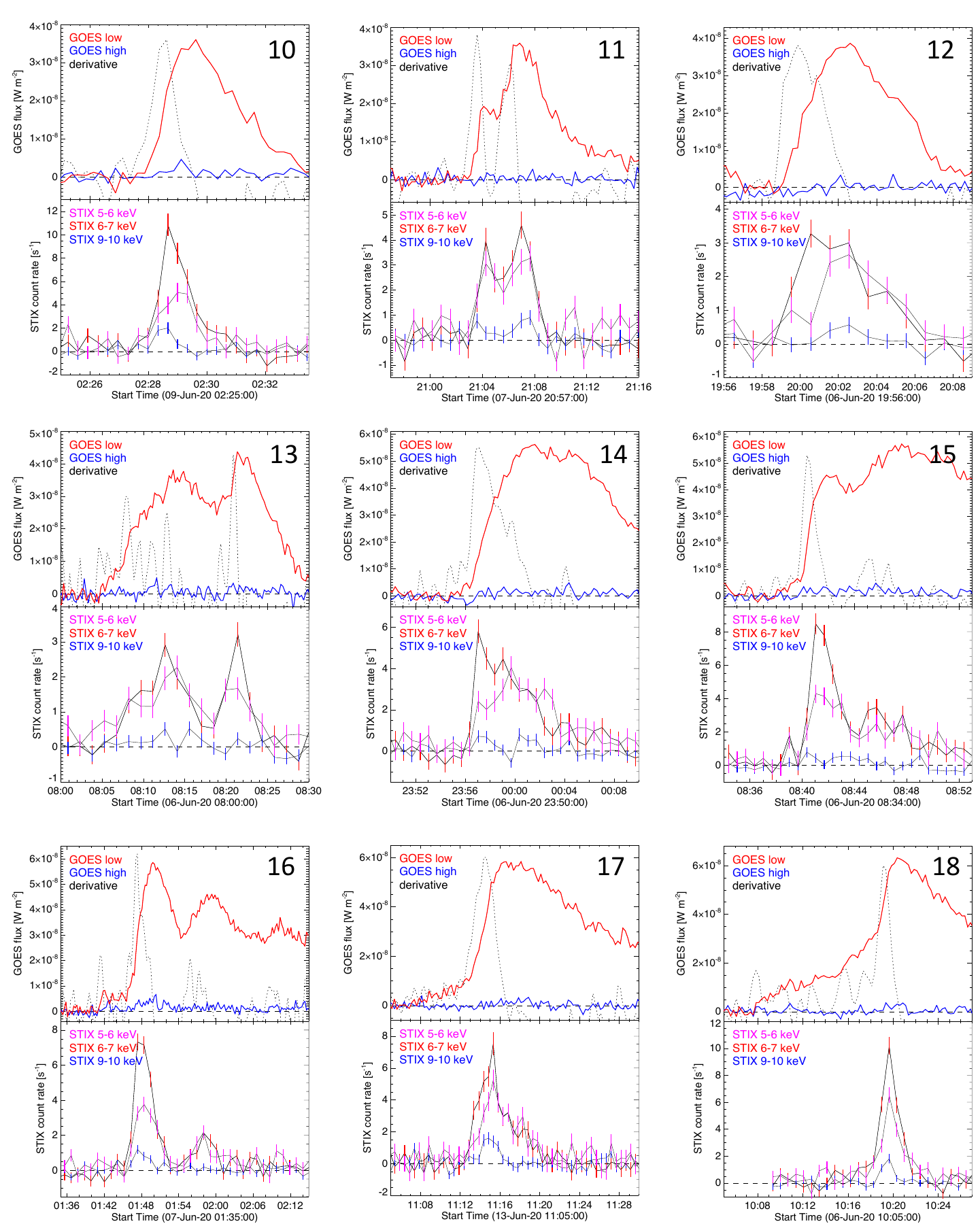}
\caption{GOES and STIX light curves for the medium size microflares (GOES A3 to A6) in our sample. Same format as previous figure.  }
\label{Fig:lc_medium}
\end{figure*}

\begin{figure*}[tbp]
\centering 
\includegraphics[height=1.25\textwidth,keepaspectratio,trim=15 20 290 00, clip]{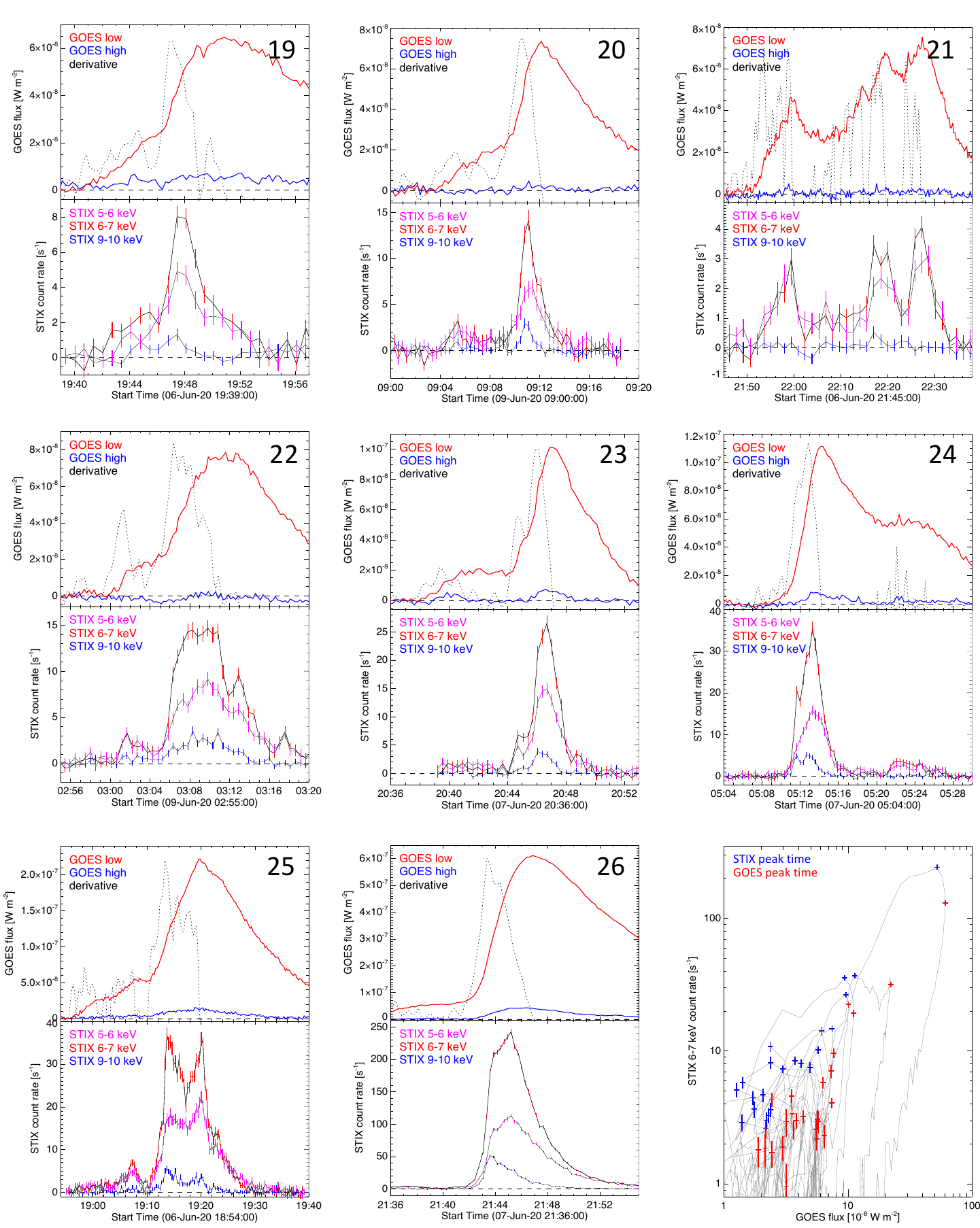}
\caption{GOES and STIX light curves for the largest size microflares (GOES A6 to B6) in our sample. Same format as previous figure. In addition, the last panel shows the scatter plot of the STIX 6-7 keV count rate as a function of the GOES 1--8 {\AA} low energy channel for all events. The blue and red points mark the values at the STIX and GOES peak time, respectively, while gray lines connect the different points according to the time evolution. 
}
\label{Fig:lc_large}
\end{figure*}

Figures 3 through 5 show background subtracted and livetime-corrected STIX count rates for three energy channels, together with the GOES light curves. The plots are ordered by the GOES class of the analyzed microflares from A2 to B6. As these are all small events, all time profiles are shown in linear scale. The displayed STIX energy bins (5-6, 6-7 and 9-10 keV) represent the three energy channels with the lowest background emission (see Fig.\,\ref{Fig:sim}). As expected from the simulation (shown in Sect.\,2), the 6-7 keV energy bin has the highest count rate with rapidly decreasing rates for bins at lower and higher energies. The 6-7 keV peak time is generally during the GOES impulsive phase (defined as rise phase of GOES low energy channel and outlined in each plot by the GOES derivative shown as dotted curve) and the duration of the STIX emissions are shorter. Clear examples for an earlier peak times in STIX profiles are flare 2, 10, and 20. There are only two events (9, 12) which show similar time profiles as GOES, but for all the others the GOES flux decays much more slowly. For a few events (e.g., 7, 17, 25, 26), the decaying GOES emission appears to also be detected in STIX, but for a shorter time. For other events (e.g., 18, 20, 23), the decay as defined by GOES does not produce a signal in STIX. 

The longer decay time of the GOES light curves is very likely associated with the GOES's  sensitivity to temperatures below STIX's temperature range (i.e., below $\sim$8 MK). As flare loops cool down, they eventually get too cold to be seen by STIX, but they are still detectable by GOES. In other words, GOES observes down to lower X-ray energies than STIX. Therefore, GOES is sensitive to lower temperature plasma than STIX is and, consequently, GOES detects the flare decay phase for longer. 

The interpretation of the earlier peak time of the STIX profiles is not as straightforward to explain because there are two effects that contribute to this phenomenon. STIX light curves are expected to peak earlier because STIX has a higher sensitivity than GOES to the hottest plasma, and the highest plasma temperature tends to be observed early in the flare time evolution \citep[e.g.,][]{2016A&A...588A.115W}. However, the earlier peak could also be attributed to the existence of a nonthermal component produced by accelerated electrons. Nonthermal signatures are most frequently observed during the the impulsive flare phase, corresponding to the GOES soft X-ray (SXR) rise time \citep{2011SSRv..159...19F}. In large flares, which tend to be hotter than microflares \citep[e.g.,][]{2007SoPh..246..339S,2005A&A...439..737B,2016A&A...588A.115W}, the 6-7 keV emission is generally completely dominated by thermal emissions. For smaller flares with lower temperatures, the associated thermal emission in X-rays falls off faster than for hotter flares, making it easier to detect nonthermal emission at lower energies. It has, indeed, occasionally been observed that microflares display nonthermal emission down to low energies \citep[e.g.,][]{2008A&A...481L..45H, 2020ApJ...891L..34G}. Hence, the earlier peak of the 6-7 keV channel cannot be attributed to one of these explanations without considering further arguments besides the timing alone. For a NuSTAR microflare, \citet{2020ApJ...891L..34G} showed that the observed Fe line complex around 6.7 keV is inconsistent with the underlying continuum spectra assuming the emission to be purely thermal, indicating that a significant fraction of the emission around the iron line complex is nonthermal. With STIX's limited spectral resolution and steep transmission reduction around and below the Fe line complex, an Fe line analysis as done by \citet{2020ApJ...891L..34G} is much more difficult to perform with STIX, in particular, in this early state of our calibration efforts. However, a detailed spectral fitting could potentially give further insights, at least for events for which a clear nonthermal component is detected at higher energies. We further investigate the question on thermal versus nonthermal emission in Sect.\,3.4, where we add EUV and soft X-ray observations as additional diagnostic tools for three selected microflares.

The bottom right panel in Fig.\,\ref{Fig:lc_large} shows, for all events, the scatter plot of the STIX 6-7 keV count rate as a function of the GOES 1--8 {\AA} flux, where points corresponding to the same event are connected by grey lines according to the time evolution. The blue points outline the values at the STIX peak time, while the red points are for the GOES peak time values. For the most significant events, the individual time evolution can be traced clearly, showing the earlier peak of the STIX light curve (blue points above red points), while the decay in STIX is much faster than for GOES. As expected, the two quantities are related to each other. As the range of flares is limited and only one flare is seen above the GOES B2 class, a fit to the data will only be published after further joint observations are recorded.

\begin{figure}[tbp]
\centering 
\includegraphics[height=0.41\textwidth,keepaspectratio,trim=250 170 230 170, clip]{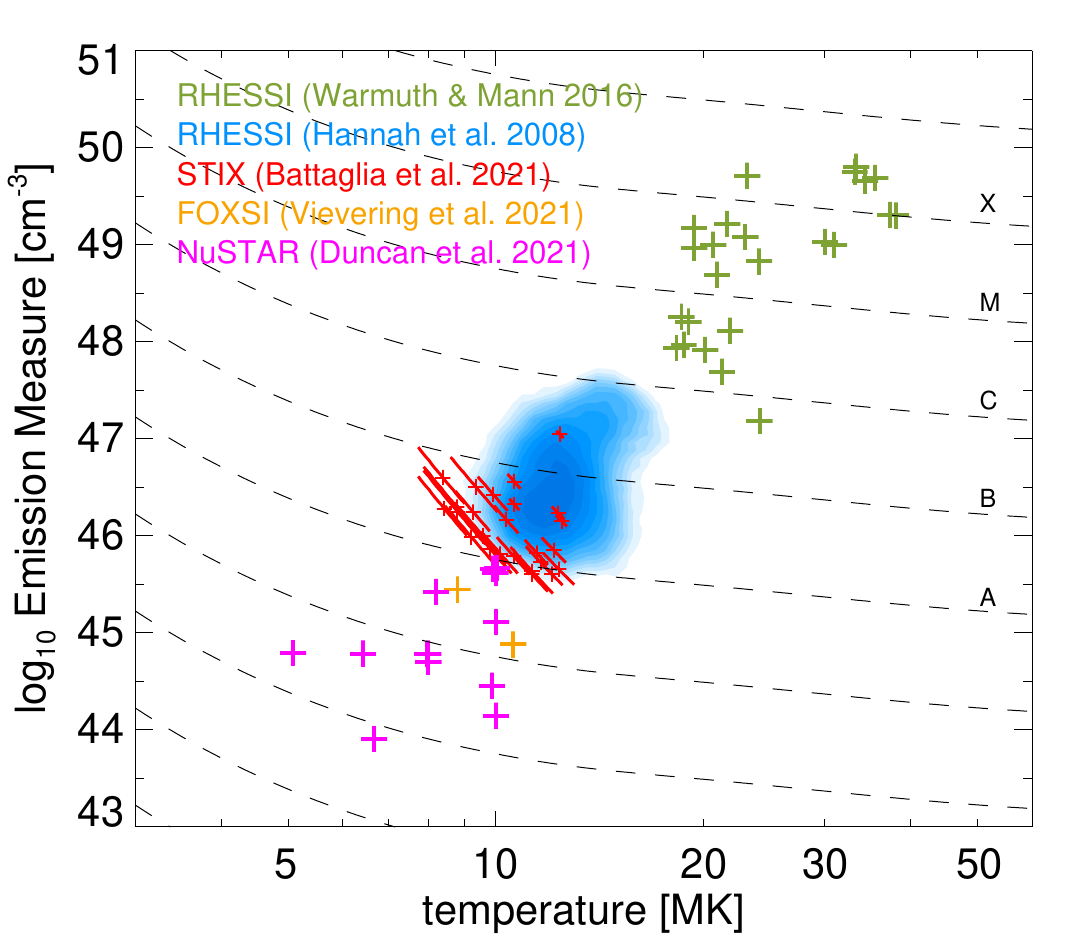}
\caption{Comparison of the emission measures and temperatures of STIX microflares detected during the commissioning phase with previously published microflare and flare observations in X-rays above 2.5 keV. The black dashed curves mark flares at equal GOES class as labeled. }
\label{Fig:xray_em_T}
\end{figure}

\begin{figure*}[tbp]
\sidecaption
\includegraphics{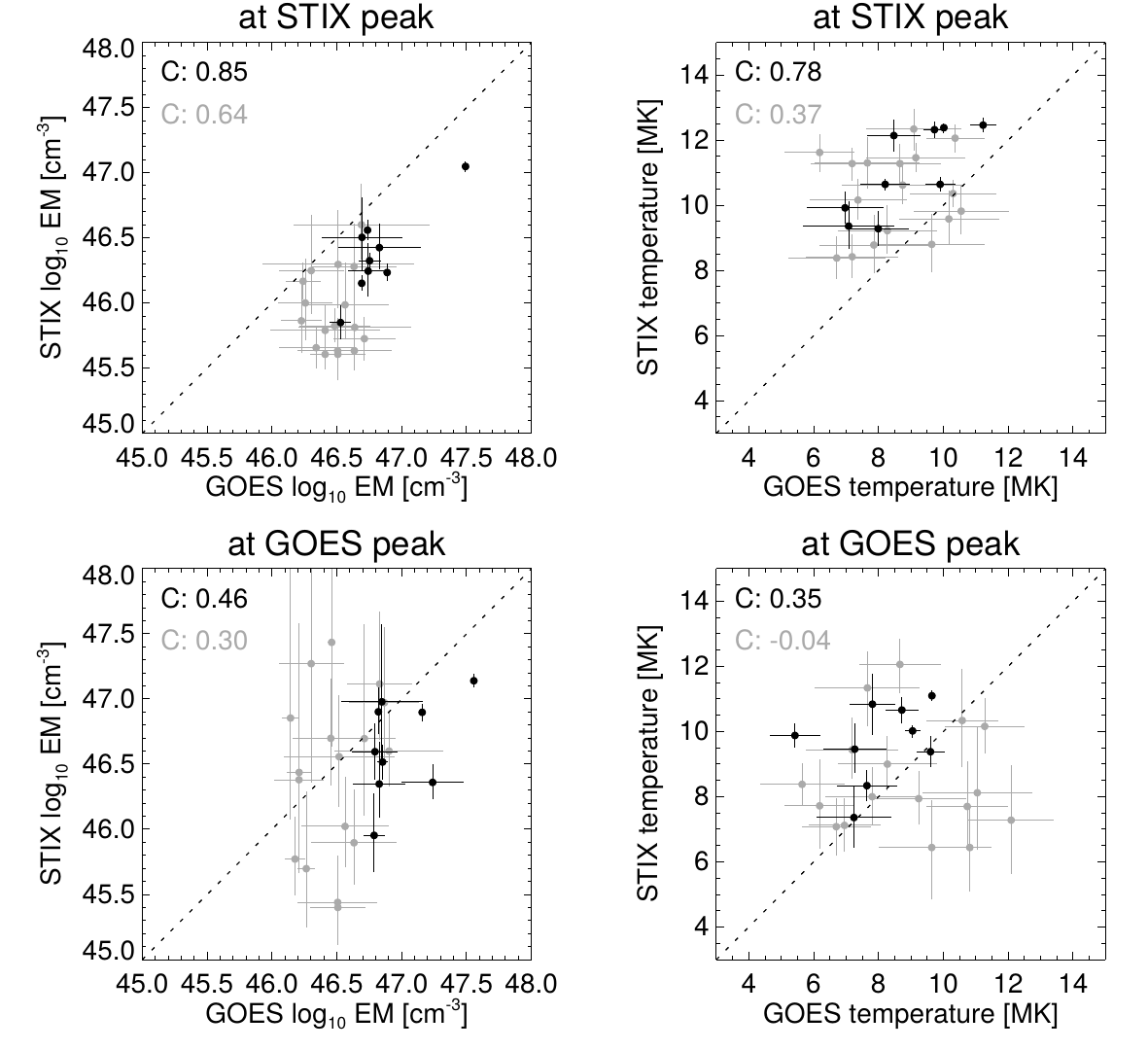}
\caption{Correlation plots of the emission measures (\textit{left}) and temperatures (\textit{right}) derived from STIX and GOES under the assumption of an isothermal plasma. The parameters were determined at the times of the peak of the STIX count rate (\textit{top}) and at the GOES peak flux (\textit{bottom}). Black symbols refer to flares with GOES class A6 or larger, while light grey symbols refer to weaker events. $C$ denotes the linear correlation coefficient for the temperatures and the logarithms of the EM (black for events larger than A6, grey for all events). The dotted lines denote $x = y$. We note that STIX generally yields lower EMs and higher temperatures than GOES during the earlier phase around the STIX peak. For the stronger events, this is also the case in the later phase around the GOES peak.}
\label{Fig:goes_stix_tem}
\end{figure*}

\subsection{Statistical temperature analysis}
\label{sec:statistical_temperature_analysis}

The results of our simple isothermal flare temperature estimates at the STIX peak time using the ratio between the 6-7 keV and 5-6 keV STIX channels is shown in Fig.\,6, together with previously published solar flare temperatures obtained via X-ray diagnostics of various instruments. Except for the five most significant events, the STIX flares during commissioning correspond roughly to the lower end of the RHESSI microflare distribution published by \citet{2008ApJ...677..704H}. Microflare observations from NuSTAR \citep[e.g.,][]{2020arXiv201106651D} and FOXSI \citep{2020arXiv201104753V} represent significantly smaller events by two orders of magnitude in emission measure, highlighting the advantages of hard X-ray focusing optics for solar flare observations, as X-rays are concentrated on small pixels with low background. There is a small range of overlap between the NuSTAR and STIX sensitivity. As a rough guideline, the largest microflares that NuSTAR can still detect before the incoming photon flux overwhelms the instrument electronics can also be detected by STIX. We note that flares 11 and 19 are actually also observed by NuSTAR. The statistical analysis also reveals that temperatures down to $\sim$8 MK can be detected with STIX, at least for microflares. Lower temperatures could potentially be detected by STIX for long-time integrated spectra in the late decay phase of large flares \citep[e.g.,][]{2002SoPh..210..341G} or for hour-long integration of active regions during non-flaring times \citep[e.g.,][]{2019ApJ...876..111I}. While these observations were taken at 0.52 AU, STIX will gain about a factor of two in sensitivity at perihelia and will therefore be able to detect slightly smaller events at those times.

What is striking about the scatter plot of flare temperatures and emission measures is the overall correlation that higher temperatures are only seen in larger flares (i.e., flares with larger emission measures). This is a strong result as X-ray diagnostics are biased towards the hottest temperatures. Hence, the absence of microflares at high temperatures is a very solid result. To illustrate this, we estimated the minimal emission measure that would be needed to see a microflare at 20 MK with STIX by repeating the simulation presented in Sect.\,2. To create a 6-7 keV count rate of the same order as the smallest flares in our sample of 26 flares (i.e., for a count rate 3 $\mathrm{s}^{-1}$), an emission measure around 10$^{44}$ cm$^{-3}$ is sufficient. Using NuSTAR for such an estimate gives an even stronger upper limit. The absence of such flares indicate that for the flare process to reach high temperatures, it also needs to create a large amount of heated plasma to account for large emission measures at high temperatures. 
In other words, a lot of energy needs to be released in order for the highest temperatures to be reached.

We now compare the emission measures and temperatures that were obtained from STIX to those that were derived from GOES. As the signal levels are comparatively low in our events, we have derived EM and temperatures from GOES fluxes integrated over one minute timesteps. Figure~\ref{Fig:goes_stix_tem} shows scatter plots of these parameters for two time periods: for the peak time of the STIX count rate (\textit{top}) and for the time of the GOES peak flux (\textit{bottom}), which occurs after the STIX peak in all events. The comparison shows that STIX generally yields smaller EMs and higher temperatures than GOES. This tendency is more pronounced at the STIX peak time, and is more clearly seen in the stronger events, namely, for GOES classes of A6 and above (highlighted in black in Fig.~\ref{Fig:goes_stix_tem}). This is also reflected in the correlation coefficients, which are significantly higher at the STIX peak and for the larger events. We went on to redo our analysis using GOES-17 data and found the same qualitative behavior.

Similar systematic differences have been found when comparing EMs and temperatures from GOES and RHESSI \citep[e.g.,][]{2005A&A...439..737B,2014SoPh..289.2547R} . These are commonly interpreted as a result of the multi-thermality of the plasma combined with the different temperature response functions of the various instruments. While GOES has a very broad temperature response (extending down to $\sim$4~MK), the response of RHESSI is weighted more strongly towards higher temperatures. Thus, the high-energy instrument will tend to give lower EMs and higher temperatures than GOES, and this evidently also applies to STIX.

Interestingly, we find that the relation between the STIX- and GOES-derived parameters changes during flare evolution. At the peak of the STIX count rate, which occurs earlier in the events, the discrepancy between the STIX- and GOES-derived parameters is larger than later in the flares when the GOES flux peaks. At the earlier time, the median STIX-derived EM amounts to only 32\%  of the GOES EM, while the STIX temperature is 2.3~MK higher, while later the median STIX EM amounts to 63\% of the GOES value, and the temperatures are roughly equal. This is consistent with the behavior reported by \citet{2016A&A...588A.115W} for a sample of C to X-class flares observed with RHESSI. It represents clear evidence that the differential emission measure (DEM) distribution, namely, the relative amount of plasma at a certain temperature in a multi-thermal plasma, is evolving as flares progress, and our STIX observations show for the first time that this is also the case in microflares. A possible physical interpretation was suggested by \citet{2016A&A...588A.115W} in terms of a combination of a cooler evaporated plasma component that is detected by both RHESSI and GOES, and a hotter directly heated coronal component that contributes significantly to the RHESSI flux, but not to the GOES flux. The hotter component is clearly observable only as long as the evaporated plasma does not dominate the emission. This interpretation could also be valid for the microflares observed with STIX, with the difference that in microflares also the hotter component will be seen by GOES, but folding its signal through the instrument response will result in a smaller contribution to the GOES result, as compared to STIX.


\subsection{Multi-wavelength analysis of three microflares \label{sec:multi-wavelength_analysis}}

The aim of this section is to place the STIX X-ray observations in the context of the multi-wavelength analysis of microflares observed together with Earth-orbiting observatories. This is not intended to be a detailed microflare study, but a first comparison of the STIX observations with the  flare morphology and evolution as revealed in complementary multi-band EUV imagery. 
In the light of the ongoing development of the ground-software and understanding of the data calibration, the detailed investigation of the microflares considered in the present section shall become the subject of a subsequent study. 

We restricted our current study to three events that are representative of the whole sample: the largest flare of our sample, along with a medium and a small size microflare, respectively. These flares originated from the same active region AR 12765 and they correspond to event 26 of Fig.\,\ref{Fig:lc_large}, event 17 of Fig.\,\ref{Fig:lc_medium}, and event 2 of Fig.\,\ref{Fig:lc_small}, with GOES classes B6, A6, and A2, respectively.
Additionally, the choice of the GOES A6 microflare was also justified by the availability of simultaneous observations by the \textit{Hinode} X-ray Telescope \citep[XRT;][]{2007SoPh..243...63G}.

It is worth mentioning the different view point of STIX with respect to
Earth-orbiting spacecrafts during the observation of the microflares. Indeed, Solar Orbiter was roughly at a distance of 0.52 AU from the Sun with a separation angle to the Sun-Earth line of about 45$^\circ$ west on June 7 and 57$^\circ$ on June 13. As seen from Earth, the active region was located south-east with respect to solar disk center when the GOES B6 and A2 microflares occurred. Six days later, when the GOES A6 microflare occurred, the active region was located south-west. As seen from Solar Orbiter, the active region was located at $\mathrm{S}800^{\prime\prime}~\mathrm{E}1600^{\prime\prime}$ on June 7, namely, close to the eastern limb and at $\mathrm{S}1000^{\prime\prime}~\mathrm{E}300^{\prime\prime}$ on June 13.

Both GOES and STIX light curves presented in this analysis have been obtained using the procedure outlined in Sects.\,\ref{sec:GOESdata} and \ref{sec:STIXdata}, respectively. The other data sets for the selected events include images and light curves of the Atmospheric Imaging Assembly \citep[AIA;][]{2012SoPh..275...17L} instrument onboard the \textit{Solar Dynamics Observatory} \citep[SDO;][]{2012SoPh..275....3P}. The AIA images shown in this section are extracted from full-disk data and the size of the field of view was chosen to be the same for all events. To process them, we adopted the standard SolarSoft programs to calibrate the data (aia\_prep) and to remove the effects of solar rotation. Consequently, the light curves have been obtained via the spatial integration of the flaring regions. 

In this analysis, we considered the $94$\,\AA, $131$\,\AA, $171$\,\AA,{} and $1600$\,\AA{} passbands at a cadence of $12$\,s for the former three and of $24$\,s for the latter. 
However, given that the AIA $94$\,\AA{} filter captures flaring emissions at the \ion{Fe}{xviii} complex as well as from lower temperatures, we isolated the \ion{Fe}{xviii} line complex \citep[see][]{2013A&A...558A..73D} to grasp emissions from higher temperatures only.
This set of AIA bands should highlight two phases of the event: the impulsive phase emission from the chromosphere (e.g. $1600$\,\AA{} and $171$\,\AA{}) and the peak time of the flare that shows mostly hot coronal flare loops (e.g. \ion{Fe}{xviii}). Depending on the microflare temperature, the $94$\,\AA{} and $131$\,\AA{} filter can have contributions from both the flare ribbons and loops. In the case of the June 13 event, XRT data of the \textit{Hinode} spacecraft are used to obtain a SXR (Al\_poly) light curve. Similarly to the AIA analysis, the standard SolarSoft xrt\_prep routine was employed to calibrate the data and solar rotation effects have been removed before light curves are derived by integrating the flaring region.

\subsubsection{Spatial evolution and morphology}

\begin{figure*}[tbp]
\centering 
\includegraphics[width=1\textwidth]{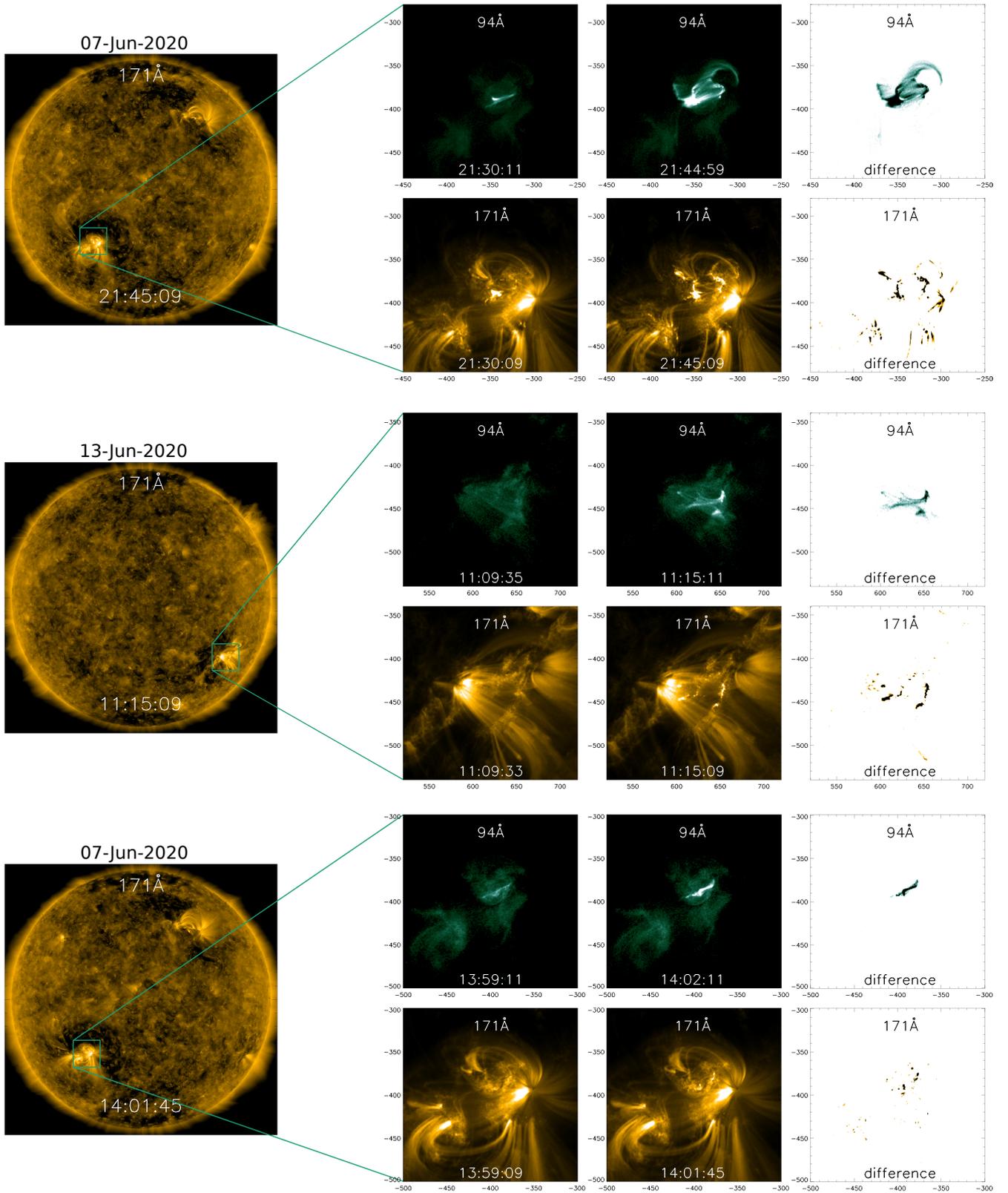}
\caption{Sequences of SDO/AIA images in the $94$\,\AA\, and $171$\,\AA\, passbands for three microflares: (\emph{top}) the B6 GOES microflare on June 7, (\emph{middle}) the A6 GOES microflare on June 13 and (\emph{bottom}) the A2 GOES microflare on June 7. Each panel is organized in the same way: on the left, a full-disk image of the $171$\,\AA{} band is shown; on the right, close-ups of the corresponding sequences of the microflares, in which the first two columns refer to distinct times while the last column indicates the difference between them. The color map of the difference image has been reversed in order to better highlight the changes due to the flares.
}
\label{Fig:maps_3events}
\end{figure*}

Before introducing the light curves of the different instruments, we present a short description of the morphologies of the microflares under study. In this paper, this is done uniquely through AIA imaging, since the calibration of the STIX imaging system has not yet concluded. 

Figure\,\ref{Fig:maps_3events} shows sequences of AIA sub-fields and a full-disk map for the three selected microflares, the GOES B6, A6, and A2 from top to bottom, respectively. Each event is portrayed in the same manner: a full-disk image in the $171$\,\AA{} passband is shown on the left, and close-ups of the corresponding microflare in the $94$\,\AA{} and $171$\,\AA{} passbands are presented on the right, for two different times. To capture the complexity within the microflares themselves, the last column on the right depicts the difference between the two timings in the first and the second column. This helps in discarding the emission from the surrounding active region, underlying the spatial and temporal changes occurring during the events.
The selected AIA images show similar features for all three events: the $171$\,\AA{} passband indicates the chromospheric emissions of the microflare ribbons, while the $94$\,\AA{} filter displays emissions from the coronal loops as well as contributions from the ribbons. The general picture arising from these images is one of a relatively complicated morphology \citep[e.g.,][]{2007SoPh..246..339S,2009A&A...505..811B,2020arXiv201106651D,2020arXiv201104753V,2020ApJ...904...94S}. However, to properly understand the structure of each microflare, the analysis of several filters is needed. This goes beyond the purpose of the paper and it is left to a subsequent study. Here, we intend to give just an overview in order to discuss the light curves in the next section.

The topmost panel of Fig.\,\ref{Fig:maps_3events} shows the GOES B6 microflare. Initially, during the pre-flare phase, a heated loop is visible in the $94$\,\AA{} filter. Afterwards, two impulsive bright ribbons with sub-structures and several coronal loops appear. Most of the emission in the the $94$\,\AA{} passband comes from the side of the loops connected to the eastern ribbon. As inferred from the full-disk map, this is due to projection effects, since the viewing angle makes the loop-top appear right above the eastern footpoint. 

The GOES A6 microflare, which is depicted in the central panel of Fig.\,\ref{Fig:maps_3events}, is most likely a microflare with crossing loops. Indeed, the loops belonging to the northern sub-structure of the western ribbon ($x\approx655^{\prime\prime}, y\approx-440^{\prime\prime}$) seem to be connected to the ribbon located near the center. Conversely, the loops belonging to the southern sub-structure of the western ribbon ($x\approx650^{\prime\prime}, y\approx-450^{\prime\prime}$) are likely to cross the previous loops, despite their second anchorage being uncertain. It is not clear whether they connect with the central ribbon or whether they cross the central loops and connect to the bright feature on the north-east side of the flaring region ($x\approx600^{\prime\prime}, y\approx-425^{\prime\prime}$).

Likewise, the GOES A2 microflare, which is shown in the lowermost panel of Fig.\,\ref{Fig:maps_3events}, exhibits a morphology different from the classical single loop scenario. Indeed, fainter emissions to the side of the main loop can be witnessed in the $94$\,\AA{} maps. In the same way, the footpoints observed with the $171$\,\AA{} band show different features on each ribbon. The western ribbon is composed of two distinct sub-structures: the one on the north seems to be connected to the fainter emission shown in the $94$\,\AA\, filter, while the one on the south belongs to the main loop emission. The ribbon in the center of the map, associated with the other end of the loop, exhibits a bright kernel toward the east ($x\approx-400^{\prime\prime},y\approx-395^{\prime\prime}$), which most likely is related to the other faint emission seen to the eastern side of the main loop.

In summary, in the present section, the morphologies of the microflares under study are presented. This serves as a context for the temporal evolution analysis reported in the following section. In general, the picture arising from the AIA images is the one of a relatively complicated morphology and to gain more insights, the analysis of several filters is needed. However, this is out of the scope of this paper and it is left to a subsequent study.

\subsubsection{Temporal evolution \label{sec:temp_evolution}} 

\begin{figure}[!]
\centering 
\includegraphics[width=0.456\textwidth]{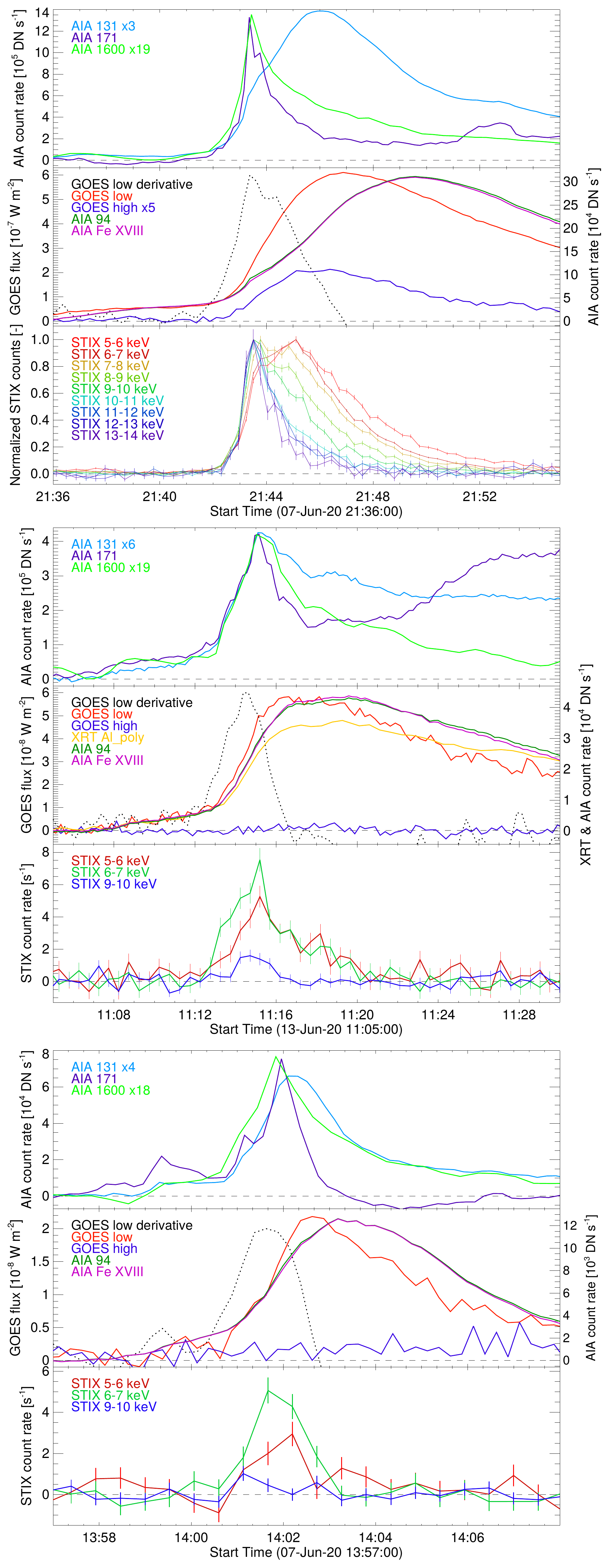}
\caption{Time profiles at various wavelengths for three microflares: (\emph{top}) the B6 GOES microflare on June 7, (\emph{middle}) the A6 GOES microflare on June 13 and (\emph{bottom}) the A2 GOES microflare on June 7. The color-coded legend in each plot indicates the nature of the light curve. Each event is illustrated as follows: in the top and  middle panels of each figure, we show the emission mostly coming from the ribbons and the coronal loops, respectively, while in the bottom panel the STIX light curves are shown.
}
\label{Fig:lc_3events}
\end{figure}

Figure \ref{Fig:lc_3events} shows the light curves of the three microflares in a selection of distinct filters for different instruments. From top to bottom, the respective time evolution of the GOES B6, A6, and A2 microflares is shown, in which every panel is organized in the same way. In the top plot, the AIA 1600 and 171\,\AA{} light curves show the impulsive phase of the events, in which the emissions mainly come from the microflares' ribbons. However, in the same plot, we added the 131\,\AA{} light curve that, depending on the microflare temperature, can show contribution from both the flare ribbons and loops. The central plot includes both GOES channels, the derivative of the GOES low energy channel, the AIA $94$\,\AA{} and the \ion{Fe}{xviii} complex, which outline the rise and decay due to the heating and cooling of the coronal loops, respectively. Finally, the bottom panel shows the STIX time profiles. In addition, for the GOES A6 event on June 13, the light curve of the Al\_poly obtained from the XRT instrument is included in the central plot. We note that for a better representation, some light curves have been scaled and the factor 
can be found in the corresponding legend.

All the considered microflares show an earlier STIX peak time compared to GOES. This may be an indication of the Neupert effect, since it corresponds to the delay between the nonthermal and thermal flare emissions \citep{1968ApJ...153L..59N,1993SoPh..146..177D,2002A&A...392..699V,2005ApJ...621..482V}. In this scenario, guided by the magnetic field lines, the previously flare-accelerated electrons travel toward the solar surface where they interact with the dense chromosphere, heat the ambient plasma and also produce nonthermal bremsstrahlung emission observed in HXRs. As a consequence of the rapid energy input, the heated chromospheric plasma expands and fills the flare loops through the process of chromospheric evaporation, producing enhanced thermal bremsstrahlung emissions at lower X-ray energies \citep[for a detailed review, see][]{2011SSRv..159...19F}.

First, let us focus on the largest event under investigation, the B6 GOES microflare on June 7 (top panel of Fig.\,\ref{Fig:lc_3events}). This flare shows an impulsive component peaking around 21:43:30 UT reaching up to 25 keV (here shown only up to 14 keV). Radio microwave observations recorded by the Expanded Owens Valley Solar Array (EOVSA) also reveals an impulsive component, with a decreasing spectra with frequency indicative of gyrosynchrotron emission (D. Gary, private communication). Hence, this early impulsive component is very likely produced by flare-accelearated electrons (the spectral analysis of this nonthermal component is presented in Sect.\,\ref{sec:spectroscopy}). The interesting feature of this microflare is the clear separation of the peak times between nonthermal and thermal emissions observed by STIX, which is not the case for the two smaller flares investigated here, as reported in subsequent paragraphs.  Indeed, the nonthermal peak time coincides with the peak times of AIA $171$\,\AA{} and $1600$\,\AA{}, consistent with the standard flare picture where accelerated electrons heat the chromosphere. In this flare, the $131$\,\AA{} filter detects also thermal emission from the coronal flare loops, since a clear delay is observable at the peak emission (about 155 s). Next, the gradual phase is highlighted in the central plot by both GOES and AIA, where a good correlation between the impulsive phase of the GOES low channel and the peak of the higher energy STIX channels can be observed. This is emphasized by the derivative of the GOES low channel, which is shown by the dotted black line. Although the peak time of the GOES 0.5-4 {\AA} high energy channel appears to be best correlated to the lower-energy STIX channels, a clear time delay can be observed between the peaks of STIX and the GOES low channel, which suggests once more the higher sensitivity of STIX toward hotter plasma, compared to GOES. Using the ratio between the STIX 6-7 and 5-6 keV channels at the peak time, the estimated single thermal temperature of the plasma is $12.4\pm0.3$\,MK with an emission measure of $11.1_{-2.0}^{+2.4} \times 10^{46}\,\mathrm{cm}^{-3}$. This is consistent with the peak time of the AIA $131$\,\AA{} shown in the upper plot. Afterwards, since GOES and AIA are sensitive to lower temperatures, we observe the plasma cooling down. Additionally, no notable differences are found between AIA $94$\,\AA{} and \ion{Fe}{xviii}, which suggests that the high-temperature component dominate the emissions in this time interval.

The tendency of earlier peak times in higher STIX energy channels is clearly discernible within this event. However, through temporal argument alone, it is not possible to quantify the thermal emission observed by the STIX lower energy channels (e.g. 5-6 keV), since there are indications that they may show part of the nonthermal bremsstrahlung emission. Indeed, both 5-6 keV and 6-7 keV channels show an impulsive increase in early times between $\sim$21:42:30\,UT and $\sim$21:43:40\,UT, which is typical of a nonthermal X-ray bremsstrahlung emission, as well as gradual increase typical of thermal emission later on, between $\sim$21:43:40\,UT and $\sim$21:45:20\,UT. With NuSTAR observations, occurrence of nonthermal bremsstrahlung emission to energies below 7 keV has already been observed \citep{2020ApJ...891L..34G}. Further progress in the ground-software and STIX imaging is needed in order to be able to answer this question.

The earlier peak times in higher STIX energy channels is less obvious during the A6 microflare, whose light curves are depicted in the central panel of Fig.\,\ref{Fig:lc_3events}. Imprints of the Neupert effect are discernible from other instruments: the impulsive manifestation is detected with AIA in the first plot, while in the second plot the gradual phase is detected by the GOES low, the SXR light curve of the XRT and both $94$\,\AA{} and \ion{Fe}{xviii}. However, it is not clear how both thermal and nonthermal emissions contribute at the different STIX energy channels at the peak time. With timing analysis alone, it is not possible to answer this question conclusively.

Interestingly, despite approaching the STIX detection limits, the higher-energy light curves in the A2 microflare peak earlier than the emission at lower energies. This event is shown in the lowermost panel of Fig.\,\ref{Fig:lc_3events}. However, distinguishing between thermal and nonthermal bremsstrahlung emissions in this case is not possible. For events with such low statistics, it is generally neither possible to differentiate emissions nor to perform STIX image reconstruction.

In general, even for large-enough microflares, caution is indicated when interpreting STIX emissions by temporal analysis alone: firstly, nonthermal emission can contribute to the observed emission even at the lowest energies, as suggested in the 5-6 keV and 6-7 keV light curves of the B6 event; secondly, if thermal bremsstrahlung would manifest earlier, then it would be impossible to distinguish between thermal and nonthermal emissions. Both STIX imaging and spectroscopy would be needed.
Hence, timing analysis alone should not be used to conclude that emission is nonthermal. The relative shifts in peak times at different wavelengths show the multi-temperature nature of the different events: a thorough differential emission measure analysis is needed to better characterize the energetics of solar flares and this will be the topic of subsequent studies. 

\subsubsection{Preliminary spectroscopic results \label{sec:spectroscopy}}

\begin{figure*}[h!]
\centering 
\includegraphics[height=0.6\textwidth]{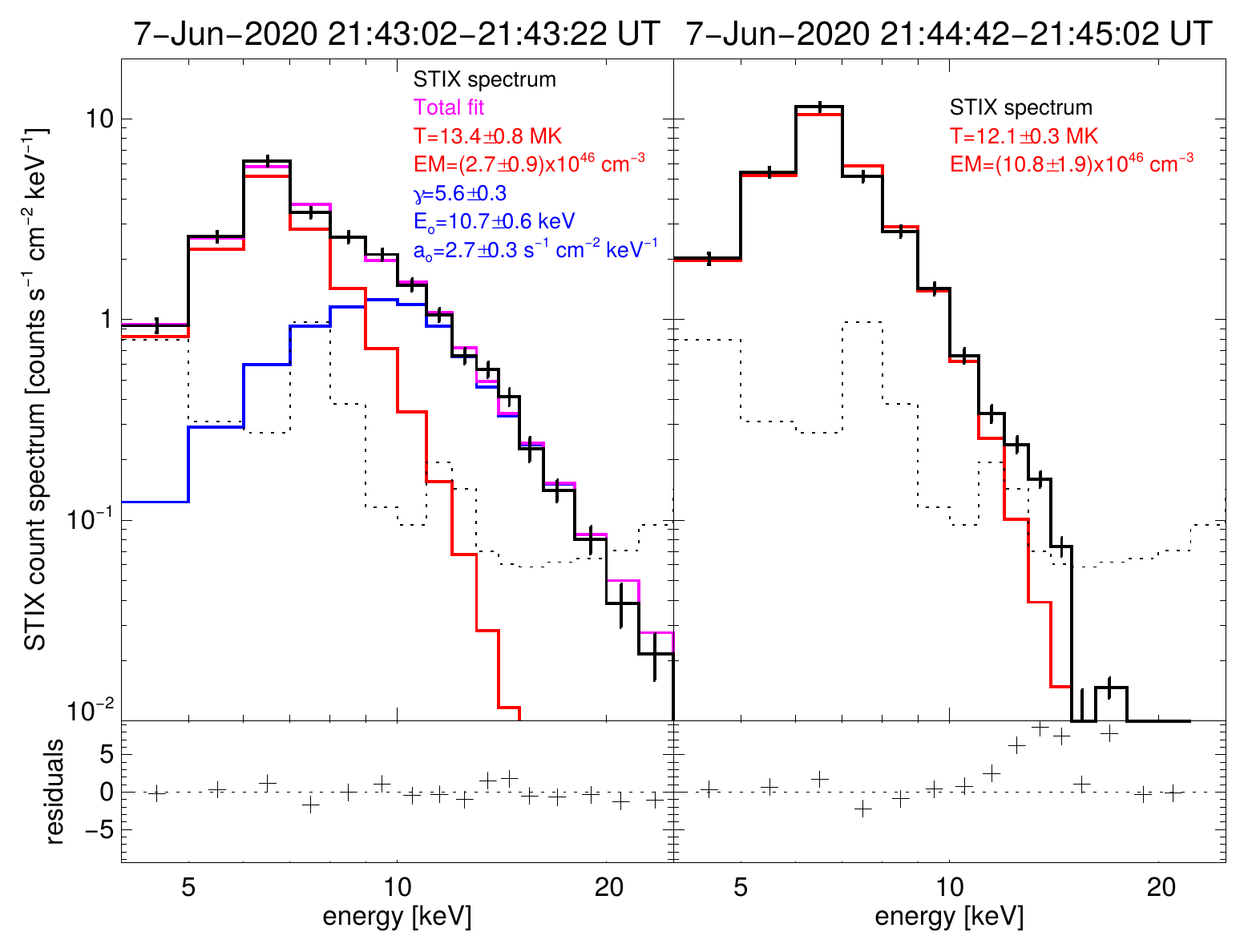}
\caption{Hard X-ray spectroscopy results of the GOES B6 microflare (flare 26) at the nonthermal (\emph{left}) and thermal (\emph{right}) peak times. The plots give the background subtracted count spectra in black, with the background shown in black dotted. The colored curves represent the fitted components: thermal in red and nonthermal power-law in blue, while the sum between them in magenta. The bottom panels show the residuals (observations minus fit) in units of the standard deviation calculated from counting statistics and a systematic error of 5\% has been added in quadrature to reflect the early stage of the calibration. The impulsive phase (\emph{left}) is fitted with both, a thermal and nonthermal component, and the combined fit is shown in magenta. The spectra around the thermal peak (\emph{right}) is only fitted by a thermal, but there is potentially still a faint trace of the nonthermal component visible above 11 keV.}
\label{Fig:spectra}
\end{figure*}

Forward fitting of solar HXR spectra is commonly done using the OSPEX\footnote{\tt{http://hesperia.gsfc.nasa.gov/ssw/packages/spex/doc/}} SSWIDL package. This software includes a wide range of commonly used functions for parametrising the thermal and nonthermal components as well as an interface to perform the fits.

For the spectroscopic analysis, the main input is a spectrogram of count rate data corrected for livetime. A Detector Response Matrix (DRM), which encapsulates the conversion of received photons to measured counts, including effects such as window transmission, is also needed. The STIX science data includes several appropriate data formats including the L1 ‘pixel’ and L4 ‘spectrogram’ data \citep{2020A&A...642A..15K}. The L1 "pixel" data allows more detailed corrections to be applied for individual pixels and detectors, such as the differing grid transmission between sub-collimators and the ELUT correction. This data also has a smaller compression error so overall it is preferable for fitting when available, but the L4 "spectrogram" data are far more compact. It also provides full spectroscopic information. While the software for routine OSPEX analysis of STIX data are currently still under construction, it is possible to convert STIX spectrogram data to a format that can be read by OSPEX. This software will be available as part of the STIX IDL ground software before the start of Solar Orbiter's nominal mission in December 2021.

For many of the smaller flares seen during commissioning the flare emission is only seen in a small number of energy channels, so a detailed spectral fit is not possible and, thus, the ratio temperature diagnostic detailed in Sect.\,\ref{sec:statistical_temperature_analysis} is appropriate. However, the B6 microflare contains significant counts over many time intervals so that spectral fits can be performed and compared. For the spectral fits shown in this section, "pixel data" \citep[large pixels only; see][]{2020A&A...642A..15K} with a 20\,s cadence for the time range from 21:37:08 to 21:52:08 UT was downloaded and analysed. As the detectors with the finest grids exhibit significantly more complex transmission and additionally have an extra Kapton covering, for simplicity, these detectors are excluded from the current analysis as the calibration of this transmission is still in progress. The Background (BKG) and Coarse Flare Locator (CFL) detectors are also excluded. The grid transmission also has a dependence on source location. An initial estimate of this, based on the coarse flare location, is applied here.

Background subtraction can be performed either within OSPEX by selecting relevant time intervals in the current spectrogram with the possibility to select time profiles which account for background variation over the flare period. The second option is to manually subtract the background from the spectrogram before passing it to OSPEX. As STIX background is dominated by the onboard calibration spectrum and it is stable over a timescale of many hours, the approach of taking a long duration background observation within a few hours of the flare provides a good estimate. Here a background observation was taken for approximately 90 minutes between 7 June 22:49 UT and 8 June 00:19 UT. These background counts undergo the same corrections as the flare observation and are subtracted directly before the spectrogram is passed to OSPEX. 

The spectra are fitted with a standard isothermal function \texttt{‘vth’} in the lower energy range and, where appropriate, based on the shape of the spectrum and the presence of significant counts in higher energy bins, a nonthermal power-law component. Indeed, EOVSA microwave observations revealing a gyrosynchrotron bursts (D. Gary, private communication) justifies the addition of a nonthermal component, instead of a second, superhot thermal component.
For simplicity and consistency with previous works on HXR microflares, the broken power-law component \texttt{"bpow"} is used. To approximate a simple power-law in photon space with a single power-law index and a low energy cutoff, we fitted only the break energy of the photon spectrum and the power-law slope above the break, fixing the value of the power-law index below the break to 1.5. 
Figure\,\ref{Fig:spectra} shows the fitted spectra for two time intervals: the peak in the higher energy channels above 10 keV (21:42:42 UT - 21:43:02 UT, defined as the peak time of the summed energy channels from 10 to 25 keV) and the time of the thermal peak (21:44:22 UT - 21:44:42 UT, defined as the peak time of the summed energy channels from 4 to 10 keV). The impulsive time bin is fit with both thermal and nonthermal components resulting in a temperature of $13.4 \pm 0.8$ MK with an EM of $(2.7 \pm 0.9$) $\times10^{46}$ cm$^{-3}$ for the thermal component and a spectral index $\gamma = 5.6 \pm 0.3 $ with cutoff energy $E_0=10.7 \pm 0.6$ keV for the nonthermal component. These fit results indicate that counts detected above $\sim$10 keV are mainly nonthermal (blue curve in Fig.\,\ref{Fig:spectra}). For completeness, we also mention the fit results for the standard cold thick target fit (\texttt{‘thick2’}). The thermal and nonthermal components obtained in this case result in a similar temperature and EM as in the previous case, but with an electron spectral index $\delta = 7.0 \pm 0.5$ and a low-energy cutoff energy $13.9 \pm 0.9$ keV. The derived energy deposition rate for electrons above the cutoff  energy becomes $(7.8 \pm 2.1) \times 10^{26}$ erg s$^{-1}$, a typical value for a B-class flare \citep{2008ApJ...677..704H}.

The thermal peak was fitted with a single isothermal function as there were insufficient counts in the higher energy bins to find a reliable power-law component. This gives a temperature of $12.1 \pm 0.3$ MK and an EM of $(10.8 \pm 1.9)$  $\times10^{46}$ cm$^{-3}$, similar to the values estimated by the ratio method for this time interval. There is a slight excess of counts above the thermal fit in the higher energy bins for this time interval, suggesting that there may still be some nonthermal emission that cannot be adequately fit  with a power-law function. The parameters derived here match well with typical HXR microflare characteristics found by \citet{2008ApJ...677..704H} who analysed the peak in the 6-12 keV energy range and found average temperature of 12.6 MK, EM = 2.5$\times10^{46}$ cm$^{-3}$, $\gamma = 6.9$ and $E_0 = 9.0$ keV. Hence, the spectra shown here reflect a typical case of a microflare, and therefore demonstrates that STIX spectroscopy provides quantitative measurements of the thermal and nonthermal component for GOES B class microflares and larger.

\subsubsection{Preliminary imaging results}

\begin{figure*}[h!]
\centering 
\includegraphics[width=1\textwidth]{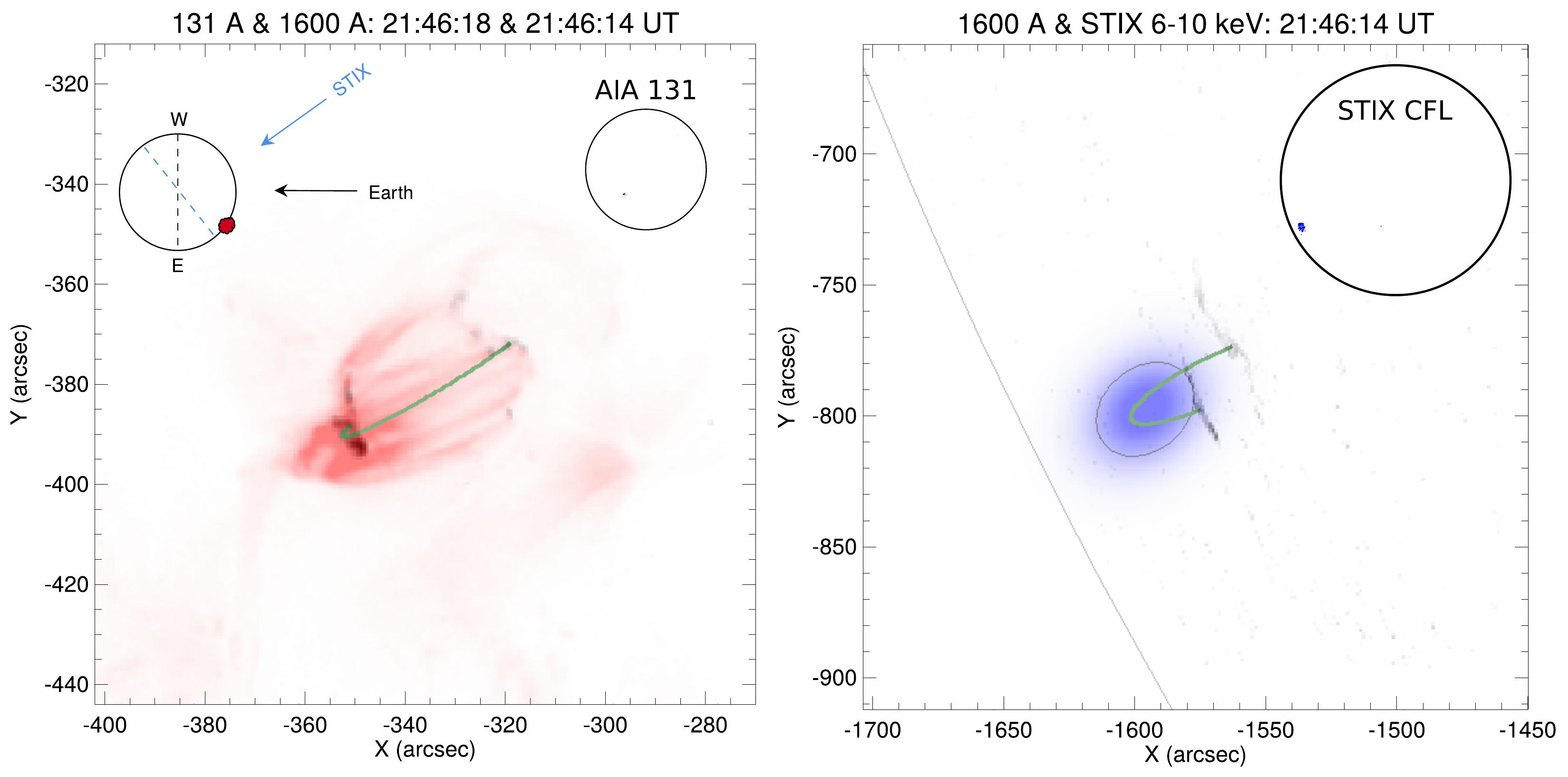}
\caption{GOES B6 microflare as seen from different vantage points. Left panel:\ Dual color-table including the background subtracted SDO/AIA $131$\,\AA{} map (red) combined with the $1600$\,\AA{} map (black) as seen from Earth, which show the flare loops and the flare ribbons, respectively. In order to guide the eye, a semi-circle perpendicular to the solar surface and connecting the flare ribbons is drawn in green. On the top-left corner, a diagram of the different point of views between Earth and STIX is outlined and on top-right the background subtracted full-disk image of the $131$\,\AA{} map, which highlights the location of the flare. Right panel:\ A dual color-table is used to show the rotated $1600$\,\AA{} map (black) and the STIX 6-10 keV image (blue) as seen from the Solar Orbiter vantage point. The green semi-circle is the same as shown in the left figure, but seen from the Solar Orbiter vantage point. In the top-right corner, the location of the flare derived by the STIX Coarse Flare Locator (CFL) is shown. As the phase calibration of the STIX imaging system is not yet completed, the absolute position of the STIX image is not yet known. For now, the STIX thermal source has been positioned to roughly agree with the top of the semi-circle. The position and size of the field of view of the two figures are chosen to show the same projected area on the Sun. The size of the solar disks shown in the inserts, on the other hand, use the same angular scale for both figures. }
\label{Fig:aia_stix_map}
\end{figure*}

STIX is an indirect imaging system that requires extensive calibration before a robust imaging pipeline can be established. The STIX team is in the middle of the calibration effort and a stable version of the imaging pipeline will be available before the start of the nominal mission. The currently available calibration of the imaging system allows us to determine simple source shapes (e.g., elliptical Gaussian), which are presented in this section. The obtained image should be seen as a preliminary step towards STIX imaging to highlight STIX’s potential, but it is not a data product that will be used in the future. The imaging analysis was done for the thermal flare source of the largest microflare in our sample (flare 26 and topmost panel of Fig.\,\ref{Fig:lc_3events}), in comparison to the AIA images.

The left panel of Fig.\,\ref{Fig:aia_stix_map} shows in red the AIA image at the peak emission of the $131$\,\AA{} band, and in black the chromospheric flare ribbons as seen by the $1600$\,\AA{} filter in a dual color-table image. As a reference, in green, a semi-circle perpendicular to the solar surface connects the two ribbons seen in the the $1600$\,\AA{} map, which gives a rough idea of how a coronal loop would appear in that location on the solar disk. On the top-left corner of the same panel, a scheme outlines the different point of views between Earth and STIX and on the top-right, a background-subtracted map of the $131$\,\AA{} shows the location of the microflare as seen from Earth. The panel to the right displays the same AIA $1600$\,\AA{} map, but rotated to take into account the different vantage point of Solar Orbiter, which corresponds to about $45$ degrees from Earth-Sun line to the west. As the flare ribbons originate from the chromosphere, we can assume that all emission originates from approximately the same known altitude, which implies that the transformation to the Solar Orbiter look direction is therefore feasible. In contrast, the $131$\,\AA{} map is not rotated, since it shows ribbons as well as coronal loops and such a transformation would largely distort the coronal source shape. However, in green, the same semi-circle is plotted in order to show how the same coronal loop would be seen from Solar Orbiter.

The STIX's CFL subcollimator provides an approximate flare location with an accuracy of around 1 arcmin, depending on counting statistics \citep[see][]{2020A&A...642A..15K}. For this paper, the CFL solution was calculated using optimized on-ground analysis software and the currently best values derived from the STIX aspect system \citep[][]{2020SoPh..295...90W}. As the aspect analysis does not yet include the calibration of the potential difference of the detector location relative to the nominal value, we added an additional uncertainty of 100$''$ to the error estimates of the derived CFL solution. The flare location of the event 26 derived from the CFL corresponds to $\mathrm{S}745\pm125^{\prime\prime}$ $\mathrm{E}1525\pm157^{\prime\prime}$ as seen from Solar Orbiter, in agreement with the expected flare location as seen by AIA (see insert within the rightmost panel of Fig.\,\ref{Fig:aia_stix_map}). 

Each STIX subcollimator provides an amplitude (the "difference between the maximum and the minimum of the Moire pattern") and phase (the "location of the peak of the Moire pattern"), which are used to reconstruct images. The calibration of the amplitudes is already available and contains a background subtraction and corrections for the grid transmission (including the flare location), ELUT and livetime (see Sect.\,\ref{sec:STIXdata}). The phase calibration, on the other hand, is more complex and it is not yet available.

To get a first STIX image from the observed amplitudes alone, we used a forward-fit algorithm that compares a pre-defined shape to the observed amplitudes \cite[see e.g.,][]{2002SoPh..210..193A}. This approach allows us to retrieve a source shape, but not a source location. Among several tested parametric models, the best fitting, namely, the lowest $\chi^2$, for the  morphology of the  event 26 is given by an elliptical Gaussian profile that is thus parameterized by its orientation, the major and minor Full Width at Half Maximum (FWHM), and the total flux. These parameters are estimated by means of a stochastic algorithm \citep[][]{488968} and the result is the blue source shown in the right panel of Fig.\,\ref{Fig:aia_stix_map}, in which the black contour represents the FWHM. As the forward fit to the amplitudes only provides the source shape, the position of the source has been selected as being the top of the semi-circle at the Solar Orbiter vantage point, since the considered integration interval occurs after the thermal peak. The flare shape is elongated, the estimated major and minor FWHM being about $35.5 \pm 5.9^{\prime\prime}$ and $27.5 \pm 4.1^{\prime\prime}$, respectively, which correspond to $13.4 \pm 2.2$~Mm and $10.4 \pm 1.5$~Mm on the Sun. The orientation, defined as the angle between the semi-major and the $x$-axis measured counterclockwise, is about $41 \pm 38$ degrees, which outlines the direction of the flare ribbons, consistent with the scenario of the presence of hot flare loops. Moreover, the source size roughly corresponds to the size of the ribbons. This suggests that the current calibration of the STIX amplitudes is performed correctly.

For completeness, we mention that the predicted total flux of the flare is $3.1 \pm 0.2$ counts s$^{-1}$ keV$^{-1}$ and the $\chi^2$ is $0.35$. The uncertainties on the parameters are estimated by performing multiple reconstructions from the data perturbed with Gaussian noise and by computing the standard deviations of the reconstructed parameters. We point out once again that this is a preliminary analysis and that a more systematic investigation of STIX imaging will be carried out by putting  more reconstruction algorithms in place and by implementing more reliable procedures for uncertainty estimation. A thorough validation of these imaging approaches will be realized once the calibrated visibility phases are ready.

\section{Summary and conclusions}\label{sec:summary}

We present the first results of the latest hard X-ray telescope to study solar flares, STIX, onboard the recently commissioned Solar Orbiter mission. Despite that calibration efforts and software developments are still ongoing, we report on 26 microflares that were observed during the Solar Orbiter commissioning phase.

First, the light curves of the three STIX energy channels with lowest non-solar background are compared to the respective GOES fluxes for all the 26 observed microflares. For all of the investigated microflares, the peak in the count spectrum is seen at the 6-7 keV science energy channel. As a rule of thumb, GOES shows a slower decay phase compared to STIX due to GOES's sensitivity to plasma at temperatures below the STIX temperature sensitivity range. Furthermore, STIX time profiles peak during the rise phase of GOES, the impulsive phase of the flare. The interpretation of this observation is ambiguous. One explanation is the higher sensitivity of STIX to hotter thermal plasma, which tends to be observed early in the flare time evolution \citep[e.g.,][]{2016A&A...588A.115W}. Another explanation is the existence of a nonthermal component produced by accelerated electrons, which are most frequently observed before the GOES peak time \citep[e.g.,][]{2002SoPh..208..297V,2011SSRv..159...19F}. Hence, despite the fact that the STIX light curves -- even at the lowest STIX science energy bins peak during the impulsive phase of the flare -- tend to mimic the Neupert effect \citep{1968ApJ...153L..59N}, these timing arguments alone should not be used to conclude that the observed emission around 6 keV is nonthermal. In addition to the time argument, the X-ray spectra should also be considered before making a definite statement. 

In a second step, the temperature and emission measure estimated for the selected set of microflares are compared to the thermal fit parameters between STIX and GOES. As most events are only seen at two STIX energy bins, a simple flux ratio method is used to derive temperatures and emission measures, instead of using the standard spectral fitting tools. Because of both multi-thermality of the plasma in solar flares and the different temperature responses, STIX generally yields smaller emission measures and higher temperatures compared to GOES (see Fig.\,\ref{Fig:goes_stix_tem}). Indeed, this effect is emphasized by the GOES temperature response that extends down to $\sim$4 MK. This is in contrast to the response of STIX that is weighted toward higher temperatures with rather limited sensitivity to plasmas below $\sim$8 MK. Hence, for any flare, STIX will always detect the hottest part of the temperature distribution, while a wider range of temperatures contribute to the broad-band GOES fluxes. The difference in the derived parameters between GOES and STIX reflect the spread of the temperature distribution. Later in the flare, when the loops are cooling, the difference is the greatest, and, therefore, GOES generally shows a longer decay time than observed by STIX. For a purely isothermal flare, the bias goes on to disappear and the derived parameters are expected to be the same by both instruments. However, we have not found such a flare, which supports the idea that flares are intrinsically multi-thermal. To account for the multi-thermal nature of solar flares, a differential emission measure analysis should be routinely applied in future thermal flare studies by combining different sets of soft and hard X-ray observations and include EUV observations as well. Although such an approach is much more labour-intensive than using the standard approach of fitting an X-ray spectrum with an isothermal distribution, it will provide a more realistic description of the thermal energy in the flare loops, which, subsequently, also allows us to better constrain the nonthermal flare emission \citep[]{2018ApJ...856L..17S}.

To compare our statistical study with previous works, STIX isothermal flare temperature estimates as a function of emission measure are compared to previous flare observations with various other instruments. The microflares observed during the commissioning phase suggest that STIX has a similar lower detection threshold to RHESSI (see Fig.\,\ref{Fig:xray_em_T}), since microflares with temperatures down to $\sim$8 MK can be detected. However, one advantage of STIX is the constant background in time compared to RHESSI's often strongly varying background, which implies a more simplified detection of small events. Moreover, while the observations in this paper are done at $0.52\,\mathrm{AU}$ from the Sun, during closest approach STIX will gain about a factor of two more in sensitivity. This will enable STIX to detect slightly smaller events during perihelia, but flares below GOES A class level as can be detected by NuSTAR and FOXSI will be generally out of reach for STIX.

In a third step, the question of whether emission observed by the lowest STIX energy channels is due to thermal or nonthermal bremsstrahlung (or a combination of both) is investigated with multi-wavelength analysis of three selected microflares. For the three selected events, the GOES B6 microflare (event 26) shows a clear nonthermal component, while it is difficult to make a definite statement for the GOES A6 and A2 flares. In the B6 flare, counts above $\sim$10 keV are found to be mainly nonthermal, and even around the Fe line complex, about $\sim$10\% of the emission is nonthermal (see Fig.\,\ref{Fig:spectra}). The peak time of the nonthermal emission nicely agrees with the peak of EUV ribbon emissions, confirming the standard flare picture in which accelerated electrons heat the chromosophere. The peak of the STIX thermal emission as defined by the 6-7 keV peak time is found about a minute and a half after the nonthermal peak, but still almost two minutes before the GOES peak. Hence, for the largest microflare, the nonthermal and thermal peaks in the STIX light curves can be clearly separated. For the two smaller flares, the counting statistics are much decreased, making such an analysis challenging, particularly at this early stage of the data analysis. However, microflares with their relatively low temperatures compared to regular flares, give us the best insight into electron acceleration down to energies below 10 keV, and future efforts will be spend to better quantify the nonthermal contributions at these lowest energies.

Lastly, first STIX spectroscopy and imaging results are shown for the largest microflare in our sample. While the calibration and data analysis tools are still under development, initial results can be already obtained, although not yet routinely by non-expert users. This first analysis demonstrates that detailed imaging and spectral analyses will be possible for microflares down to at least the GOES B class level. However, STIX images taken from a different vantage compared to the Earth-Sun line pose new challenges for comparing with images taken from Earth bound observatories. A first attempt at comparing AIA and STIX images is shown in Fig.\,\ref{Fig:aia_stix_map}.

In conclusion, we demonstrate that STIX works as designed, by analysing microflare observations taken during the Solar Orbiter commissioning phase. The presented set of microflares highlights the STIX diagnostics and gives insights into the analysis step beyond what is discussed in the STIX instrument paper. Once the STIX analysis tools are fully developed, multi-wavelength analyses in combined studies with the other Solar Orbiter instruments and Parker Solar Probe, as well as Earth-orbiting and ground-based solar telescopes will provide essential new insights in understanding the particle accelerations and energy release in solar flares.

\begin{acknowledgements}
This paper is dedicated to the memory of Richard Schwartz\footnote{https://www.forevermissed.com/richard-alan-schwartz/about}. We wish to acknowledge the referee of this paper, L. Glesener, for the valuable inputs and encouraging comments. Solar Orbiter is a space mission of international collaboration between ESA and NASA, operated by ESA. We thank all the individuals who contributed to STIX, and all the funding agencies that supported STIX: Swiss Space Office, the lead funding agency for STIX, the Polish National Science Centre (grants 2011/01/M/ST9/06096 and 2015/19/B/ST9/02826), Centre national d'études spatiales (CNES), Commissariat à l'énergie atomique et aux énergies alternatives (CEA), the Czech Ministry of Education (via the PRODEX program), Deutsches Zentrum f\"ur Luft- und Raumfahrt (DLR) (grants: 50 OT 0903, 1004, 1204), the Austrian Space Programme, ESA PRODEX, administered in Ireland by Enterprise Ireland, the Agenzia Spaziale Italiana (ASI) and the Istituto Nazionale di Astrofisica (INAF).

AFB and HX are supported by the Swiss National Science Foundation Grant 200021L\_189180 for STIX. JS and AMV acknowledge the 
Austrian Science Fund (FWF): I4555-N. LK is supported by a SNSF PRIMA grant. FF and JK acknowledge support by the project RVO:67985815.

\end{acknowledgements}

%
%

\bibliographystyle{aa} 
\bibliography{aanda} 

\end{document}